%% file: bare_jrnl.tex
\documentclass[journal]{IEEEtran}
\usepackage[backend=bibtex,style=numeric,citestyle=numeric-comp,sorting=none]{biblatex}
\usepackage{xcolor}
\usepackage{amsmath,amssymb,amsfonts}
\usepackage{algorithmic}
\usepackage{graphicx}
\usepackage{subfigure}
\usepackage{caption}                

\usepackage{array}
\usepackage{multirow}
\usepackage{textcomp}
\usepackage{hyperref}
\usepackage{adjustbox}
\usepackage{booktabs, multirow} 
 \usepackage{makecell}
\usepackage{comment}
\usepackage{tabularx}
\usepackage{soul}

\bibliography{stateofart}
\providecommand{\keywords}[1]{\textbf{\textit{Index terms---}} #1}
\begin{document}

\title{A Review of the In-Network Computing and Its Role in the Edge-Cloud Continuum}


\author{
    \IEEEauthorblockN{Manel Gherari\IEEEauthorrefmark{1}, Fatemeh Aghaali Akbari\IEEEauthorrefmark{4}, Sama Habibi\IEEEauthorrefmark{3}, Soukaina Ouledsidi Ali\IEEEauthorrefmark{1}, Zakaria Ait Hmitti\IEEEauthorrefmark{1}, Youcef Kardjadja\IEEEauthorrefmark{2}, Muhammad Saqib\IEEEauthorrefmark{1}}, Adyson Magalhaes Maia\IEEEauthorrefmark{2}, Marsa Rayani\IEEEauthorrefmark{4}, 
    Ece Gelal Soyak\IEEEauthorrefmark{5}, Halima Elbiaze\IEEEauthorrefmark{1}, Ozgur Ercetin\IEEEauthorrefmark{3}, Yacine Ghamri-Doudane\IEEEauthorrefmark{2}, Roch Glitho\IEEEauthorrefmark{4}, Wessam Ajib\IEEEauthorrefmark{1}
    
    \IEEEauthorblockA{\IEEEauthorrefmark{1} Université du Québec à Montréal, Montreal, Canada\\}
    \IEEEauthorblockA{\IEEEauthorrefmark{2} La Rochelle University, La Rochelle, France\\}
    \IEEEauthorblockA{\IEEEauthorrefmark{3} Sabanci University, Istanbul, Turkey\\}
    \IEEEauthorblockA{\IEEEauthorrefmark{4} Concordia University, Montreal, Canada\\}
    \IEEEauthorblockA{\IEEEauthorrefmark{5} Bahcesehir University, Istanbul, Turkey\\}
}

\maketitle

\begin{abstract}

Future networks are anticipated to enable exciting applications and industrial services ranging from Multisensory Extended Reality to Holographic and Haptic communication.  
These services are accompanied by high bandwidth requirements and/or require low latency and low reliability, which leads to the need for scarce and expensive resources.
Cloud and edge computing offer different functionalities to these applications that require communication, computing, and caching (3C) resources working collectively. Hence, a paradigm shift is necessary to enable the joint management of the 3Cs in the edge-cloud continuum.
We argue that In-Network Computing (INC) is the missing element that completes the edge-cloud continuum. This paper provides a detailed analysis of the driving use-cases, explores the synergy between INC and 3C, and emphasizes the crucial role of INC. A discussion on the opportunities and challenges posed by INC is held from various perspectives, including hardware implementation, architectural design, and regulatory and commercial aspects.

\end{abstract}

\keywords{In-Network Computing, Edge-Cloud Continuum, Programmable Network Device, Communication, computing, and caching integration}

\IEEEpeerreviewmaketitle


\input{Content/introduction}


 

\input{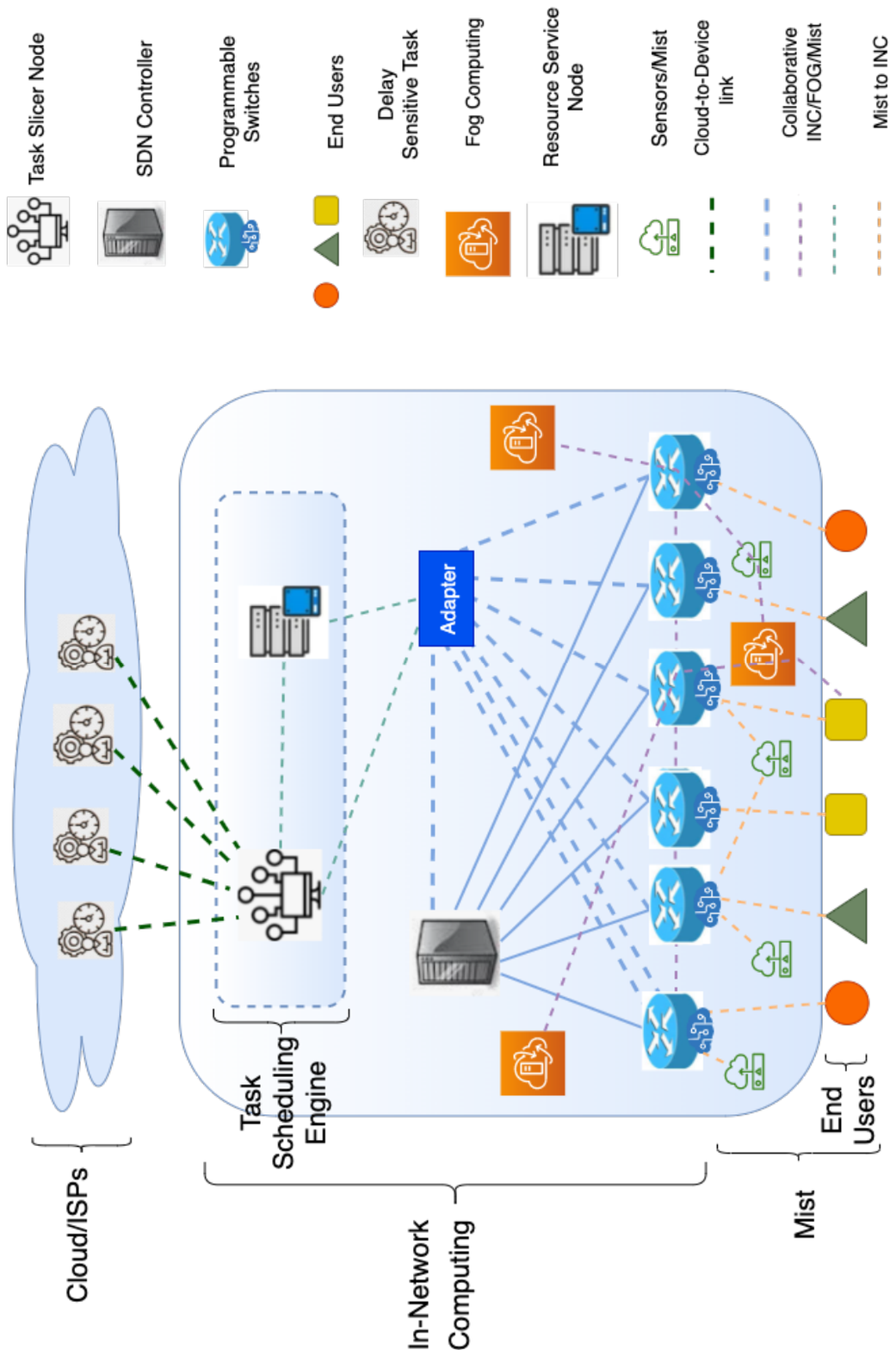}

\input{Content/challenges_opportunities_C4}

\input{Content/Conclusion}




\section*{Acknowledgment}
The authors would like to thank all SCORING project team members.

\printbibliography
\end{document}

%% file: Content/introduction.tex
\section{Introduction}

The proliferation of data, resulting from the continuous increase in IoT (Internet of Things) device deployments, has resulted in the emergence of cloud data storage and associated processing services. Centralized cloud solutions are considered insufficient for novel applications such as holographic communications with strict requirements, including low delay and high bandwidth. As a complementary solution, edge computing brings cloud services closer to users and allows different edge devices to run heterogeneous applications that use and produce all data types. Thus, prominent solutions are required for moving data, storing it, and processing it throughout the Edge-Cloud continuum and in between. Therefore, a seamless cooperation between computing, caching, and communication (3C) is essential. Recent studies (see Table \ref{tabe:RelatedWork}) have focused on this collaboration mainly at the edge, addressing the management of 3C resources and ignoring an essential factor that \textit{the network can no longer be considered as a bare data transport medium.} 

With recent advances in network virtualization and softwarization, seamless 3C collaboration will be possible at points between edge and cloud \cite{zeng2021guest}. In particular, In-Network Computing (INC) has emerged as a new paradigm in which network core infrastructures do not just transmit data but also act on it (e.g., perform computation and caching). 
INC refers to the offloading of application-specific tasks from the end-host to the programmable network devices (e.g., programmable switch, Smart Network Interface Card (NIC)). Since INC performs computing inside the network, the transaction terminates within the path, thereby avoiding unpredictable latency in the communication path. Besides benefiting from the pipeline design of network devices, INC offers higher orders of magnitude in terms of throughput processing capacity than can be achieved by an end-host server.
The rapid evolution of Programmable Network Devices (PND) such as Barefoot Tofino switches \cite {Tofino} facilitate data plane programmability to perform computer-like functions such as data aggregation, task scheduling, and traffic classification in the network.
Thus, INC is becoming another pillar of technological enablers - network computerization - towards seamless 3C collaboration in the Edge-Cloud Continuum.

This paper provides a comprehensive review of the literature examining the synergy between INC and 3Cs. We provide a systematic analysis of how current research addresses developments and contributions made in INC and builds an implicit layer supporting 3C cooperation in the cloud-Edge continuum.

\begin{table*}
\caption{Related Work Classification}\label{tabe:RelatedWork}
\centering
\begin{adjustbox}{width=1\textwidth}
\begin{tabular}{|l|l|l|l|l|l|} 

\hline
\multicolumn{1}{|c|}{\multirow{2}{*}{\textbf{Paper}}}   & \multicolumn{1}{c|}{\multirow{2}{*}{\textbf{year}}} & \multicolumn{3}{c|}{\textbf{ Content Coverage}}  &  \multicolumn{1}{c|}{\multirow{2}{*}{\textbf{Contribution}}}  \\ 
\cline{3-5}
\multicolumn{1}{|c|}{}  & \multicolumn{1}{c|}{} & \textit{\textbf{\qquad Cloud}}  & \textit{\textbf{\qquad Core}}  & \textit{\textbf{\qquad Edge}} &  \multicolumn{1}{c|}{}  \\ 
\hline

C.Wang et \textit{al.} \cite{8060515} & 2018 & \begin{tabular}{@{\labelitemi\hspace{\dimexpr\labelsep+0.5\tabcolsep}}l@{}}Computing\\Caching\end{tabular}   & \begin{tabular}{@{\labelitemi\hspace{\dimexpr\labelsep+0.5\tabcolsep}}l@{}}Caching\end{tabular} & \begin{tabular}{@{\labelitemi\hspace{\dimexpr\labelsep+0.5\tabcolsep}}l@{}}Computing\\Caching\end{tabular}         &                       
\begin{tabular}[c]{@{}l@{}}Integration of Networking, Caching, and Computing in Wireless Systems: \\A Survey, Some Research Issues, and Challenges\end{tabular}\\
\hline

S.Wang et \textit{al.} \cite{7883826}   & 2017                                                & \begin{tabular}{@{\labelitemi\hspace{\dimexpr\labelsep+0.5\tabcolsep}}l@{}}Computing\end{tabular}            & \begin{tabular}{@{\labelitemi\hspace{\dimexpr\labelsep+0.5\tabcolsep}}l@{}}Caching\end{tabular} & \begin{tabular}{@{\labelitemi\hspace{\dimexpr\labelsep+0.5\tabcolsep}}l@{}}Computing\\Caching\end{tabular}                  & \begin{tabular}[c]{@{}l@{}} A Survey on Mobile Edge Networks: \\Convergence of Computing, Caching, and Communications\end{tabular}  \\ 
\hline

M.Mehrabi et \textit{.al} \cite{8896954}       & 2019        & \begin{tabular}{@{\labelitemi\hspace{\dimexpr\labelsep+0.5\tabcolsep}}l@{}}Computing\end{tabular}      & \begin{tabular}{@{\labelitemi\hspace{\dimexpr\labelsep+0.5\tabcolsep}}l@{}}Caching\end{tabular} & \begin{tabular}{@{\labelitemi\hspace{\dimexpr\labelsep+0.5\tabcolsep}}l@{}}Computing\\Caching\end{tabular}                  &  \begin{tabular}[c]{@{}l@{}}Device-Enhanced MEC: Multi-Access Edge Computing (MEC) Aided by\\~End Device Computation and Caching: A Survey\end{tabular}    \\ 
\hline

P.Mach et~\textit{.al} \cite{7879258}      & 2017       &        &    & 
\begin{tabular}{@{\labelitemi\hspace{\dimexpr\labelsep+0.5\tabcolsep}}l@{}}Computing\\Caching\end{tabular}                  &  Mobile Edge Computing: A Survey on Architecture and Computation Offloading\\ 
\hline

X.Wang et~\textit{.al} \cite{8976180}  & 2020  &     &     & \begin{tabular}{@{\labelitemi\hspace{\dimexpr\labelsep+0.5\tabcolsep}}l@{}}Computing\\Caching\end{tabular}   &  Convergence of Edge Computing and Deep Learning: A Comprehensive Survey  \\ 
\hline

J.Ren~~\textit{.al} \cite{10.1145/3362031} & 2019    & \begin{tabular}{@{\labelitemi\hspace{\dimexpr\labelsep+0.5\tabcolsep}}l@{}}Caching\end{tabular}         &    & \begin{tabular}{@{\labelitemi\hspace{\dimexpr\labelsep+0.5\tabcolsep}}l@{}}Caching~\\Computing\end{tabular}                 &   \begin{tabular}[c]{@{}l@{}}A Survey on End-Edge-Cloud Orchestrated Network Computing Paradigms: \\Transparent Computing, Mobile Edge Computing, Fog Computing, and Cloudlet\end{tabular}  \\ 
\hline

Y.Zhao~\textit{.al} \cite{8758431}      & 2019                                                & \begin{tabular}{@{\labelitemi\hspace{\dimexpr\labelsep+0.5\tabcolsep}}l@{}}Computing\\Caching\end{tabular} &                                                                                                 & \begin{tabular}{@{\labelitemi\hspace{\dimexpr\labelsep+0.5\tabcolsep}}l@{}}Caching\end{tabular}                             & Edge Computing and Networking: A Survey on Infrastructures and Applications\\ 
\hline

NC.Luong .al \cite{8714026} & 2019   &      &   & \begin{tabular}{@{\labelitemi\hspace{\dimexpr\labelsep+0.5\tabcolsep}}l@{}}Communication\\Caching\\Computing\end{tabular} & \begin{tabular}[c]{@{}l@{}}  Applications of Deep Reinforcement Learning in Communications and Networking:\\ A Survey\end{tabular} \\ 
\hline

Q Luo~\textit{.al} \cite{9519636}\ & 2021  &    &        & \begin{tabular}{@{\labelitemi\hspace{\dimexpr\labelsep+0.5\tabcolsep}}l@{}}Communication\\Caching\\Computing\end{tabular} &  Resource Scheduling in Edge Computing: A Survey \\ 
\hline

L.Bittencourt~\textit{.al} \cite{bittencourt2018internet}  & 2018    & \begin{tabular}{@{\labelitemi\hspace{\dimexpr\labelsep+0.5\tabcolsep}}l@{}}Communication \end{tabular} & & \begin{tabular}{@{\labelitemi\hspace{\dimexpr\labelsep+0.5\tabcolsep}}l@{}}Communication\end{tabular}  &  The Internet of Things, Fog and Cloud continuum: Integration and challenge    \\ 
\hline

B.Sonkoly~\textit{.al} \cite{placementSurvey} & 2021    &   &    & \begin{tabular}{@{\labelitemi\hspace{\dimexpr\labelsep+0.5\tabcolsep}}l@{}}Caching\\Computing\end{tabular} & \begin{tabular}[c]{@{}l@{}}Survey on Placement Methods in the Edge and Beyond\end{tabular} \\ 
\hline
J.Santos et~\textit{.al} \cite{9476028} & 2021   &   &    &\begin{tabular}{@{\labelitemi\hspace{\dimexpr\labelsep+0.5\tabcolsep}}l@{}} Caching \end{tabular} & 
\begin{tabular}[c]{@{}l@{}}Towards Low-Latency Service Delivery in a Continuum of Virtual Resources:\\State-of-the-Art and Research Directions\end{tabular}                              \\ 
\hline
S. Kianipseh et~\textit{.al} \cite{INC22} & 2022   &  \begin{tabular}{@{\labelitemi\hspace{\dimexpr\labelsep+0.5\tabcolsep}}l@{}} Computing\end{tabular} & \begin{tabular}{@{\labelitemi\hspace{\dimexpr\labelsep+0.5\tabcolsep}}l@{}}Communication\\Caching\\Computing\end{tabular} &\begin{tabular}{@{\labelitemi\hspace{\dimexpr\labelsep+0.5\tabcolsep}}l@{}}Computing\end{tabular}  & 
\begin{tabular}[c]{@{}l@{}}A Survey on In-Network computing: Programmable Data Plane And technology specific applications\end{tabular}                              \\ 
\hline

Our Review & 2022   &   \begin{tabular}{@{\labelitemi\hspace{\dimexpr\labelsep+0.5\tabcolsep}}l@{}}Communication\\Caching\\Computing\end{tabular}   & \begin{tabular}{@{\labelitemi\hspace{\dimexpr\labelsep+0.5\tabcolsep}}l@{}}Communication\\Caching\\Computing\end{tabular}  & \begin{tabular}{@{\labelitemi\hspace{\dimexpr\labelsep+0.5\tabcolsep}}l@{}}Communication\\Caching\\Computing\end{tabular}    & \begin{tabular}[c]{@{}l@{}} A Review of the In-Network Computing and Its Role in the Edge-Cloud Continuum\end{tabular}  \\
\hline
\end{tabular}
\end{adjustbox} 
\end{table*}

Our contributions are summarized as follows.
\begin{itemize}
\item An overview of today's distributed computing landscape and related services is given.

\item An extensive description of in-network computing, including a taxonomy, technological enablers, and implementation techniques, is provided.

\item A detailed analysis of several use cases emphasizes the need for in-network computing to meet the stringent requirements of emerging applications.

\item The challenges of in-network computing are examined from various perspectives, including equipment implementation, architectural design, regulation, and business.

\item Opportunities and future directions, such as the development of autonomous networks, the integration of artificial intelligence into networks, and the addition of control aspects to 3C, are discussed.

\end{itemize}

The remainder of this paper is organized as follows. We describe the existing computing landscape in Section \ref{ComputingParadigm}. In Section \ref{INC}, we lay the foundation for the entire survey by describing the fundamental concepts of INC. Section \ref{INC_Enablers} describes key INC enablers. Service provisioning functionalities in INC environments are presented in Section \ref{ResourceAllocation}. We describe main INC implementation approaches in Section \ref{Implementation}. Finally, Section \ref{Challenges} highlights the challenges and the opportunities related to INC as well as open research directions. 

\input{Content/computing_paradigms}

%% file: Content/computing_paradigms.tex
\section{A New Era of Massively Scalable Computing:  Cloud-Edge-Fog-Dew-Mist Continuum}\label{ComputingParadigm}

\begin{figure}
    \centering
    \includegraphics[width=0.48\textwidth]{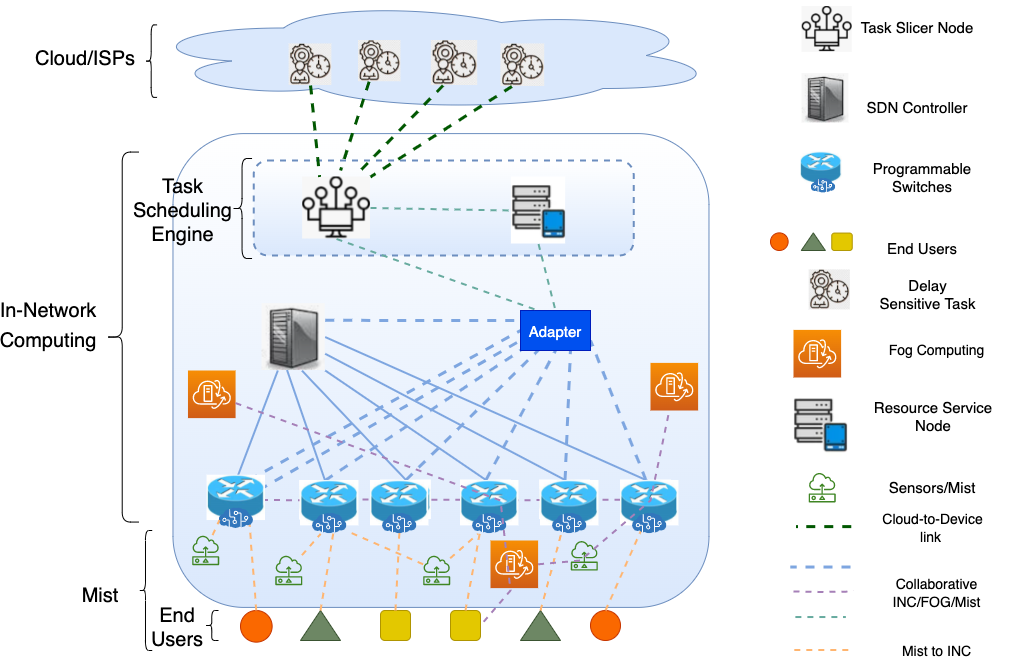}
    \caption{In-network computing supported ecosystem of networked devices}
    \label{fig:INC}
\end{figure}

The following presents a short overview of the different computing paradigms present in today's networks and discusses how INC can act as a facilitator of these paradigms.\\
The widespread use of cloud computing has reduced the need for local storage for edge devices and has enabled edge devices to delegate computationally intensive tasks. On the other hand, computing tasks should be transferred to devices near edge devices to reduce latency and improve resistance in the event of heavy traffic in the network backhaul. Thus, in recent years, the interest in edge computing has increased \cite{FIROUZI2021101840}.  One of the main advantages of 5G over earlier mobile network generations is the ability to process data generated by devices on or near mobile network edges via Multi-access Edge Computing (MEC). MEC servers are represented by a set of computing nodes purposefully installed at the network edge by a mobile network operator.  Hence, 5G can contribute greatly to latency assurance and load reduction on the network backhaul. MEC's support for caching, processing, and distributed computing can meet the needs of end-user devices from high mobility vehicles to low latency manufacturing floor machines \cite{7931566}. However, the initial installation of MEC servers may not be appropriate for all applications and may not be flexible. \textit{Fog Computing} (FC), therefore addresses this inadequacy by generally including consumer devices that provide faster installation of available resources \cite{7543455}. \textit{Dew Computing} (DC), closely related to cloud computing, is another paradigm associated with data storage security and reliability \cite{Wang2016DefinitionAC}. The aim of DC is to extend the functionality of the cloud to edge devices, where services can run both independently and in collaboration with the cloud.  
By forming adhoc clusters, groups of edge devices may mimic a larger processing platform such as a MEC server or an FC node, and process tasks offloaded to them by other edge devices \cite{894385}. While edge devices come together to process a computationally intensive task, resource-rich extreme edge devices can decide to individually provide their processing power to adjacent edge computing nodes \cite{8486685}. Such cases, where data processing is carried out by stationary extreme edge devices, are also referred to as \textit{Mist computing} \cite{lea2020iot}. Mist computing (MC) devices may exist alongside the FC architecture. MC devices are represented by a thin layer of resource-constrained computing devices, wedged between the sensing devices and the first layer of the fog architecture. 
Note that Cloud, FC, DC and MC have different functionalities that help applications achieve stringent service requirements that cannot be met unless they work together. \textit{In-network computing} (INC) has the potential to strengthen the above-mentioned computational paradigms as depicted in Figure \ref{fig:INC}. 

Distributed computing services explained above may employ serverless computing  to easily harness the power of the cloud. With serverless computing, developers simply provide an event-driven function to cloud providers, and the provider seamlessly scales function invocations to meet demands as event triggers occur. Today, successful serverless products fall into the application-specific category and are narrowly targeted, whereas general-purpose serverless abstractions have a better chance of displacing serverful computing. 
Automatic optimization with machine learning will play an important role in all service models discussed above. It may help decide where to run the code, where to keep the state, when to start up a new execution environment, and how to keep utilization high and costs low while meeting performance objectives \cite{10.1145/3406011}.

Managing the massive amount of data at the edge requires a distributed and collaborative computing hierarchy. Distributed learning (DL) and collaborative learning (CL) are two learning methods used in the cloud-fog-mist continuum.  DL and CL have been traditionally used for managing and processing a large amount of data and heavily loaded resources. DL focuses on learning mechanisms on different clients through a network system, and CL focuses on the integration of distributed learning on different clients. In particular, the emergence of edge intelligence provides DL and CL with the necessary computational power from the heterogeneous devices located at the network edge.  This optimizes the reliability of the network topology and thus achieves greater efficiency and better performance.

Meanwhile, the training of machine learning (ML) models on edge devices is difficult, especially due to their computational and memory limitations. In general, offloading tasks from a central cloud server to edge nodes results in a substantial reduction of transmitted data; however, ML models need massive computational resources, including energy and memory. 
This requirement can create bottleneck issues when ML is used with edge/fog computing. 
When resources are limited, the successful use of ML requires  an edge server to have many different implementations. Each distinct implementation of the network generates the same classification or prediction with varying accuracy versus resource in a way that a higher service level will produce greater accuracy. 

Finally, \textit{energy-driven computing} is a paradigm that empower edge devices with autonomous and low-energy operation \cite{sliper2020energy}.  Note that, the tens of billions of connected devices envisioned in the next generation of networks will require novel solutions to avoid the cost and maintenance requirements imposed by battery-powered operation. Energy harvesting is a promising technology to this end, but harvested energy typically fluctuates, often unpredictably, and with large temporal and spatial variability. Energy-driven computing allows devices to sleep through periods of no energy, endure periods of scarce energy, and capitalize on periods of ample energy. Intermittent operation should be endured since power failures can be frequent and unpredictable. However, by retaining computational progress through power cycles, intermittent computing systems allow long-running applications to progress incrementally whenever energy is available. A key consequence is QoS must often be sacrificed because the devices are not continuously operational \cite{9403911}. Hence, intermittent operation is only applicable to certain applications, e.g., for example, an autonomous solar-powered quadcopter whose mission is to collect environmental sensor data.  Although INC may not be limited by energy, it may also benefit from intermittent computing. For example, a smart NIC may preemptively interrupt an ongoing computation for a flow to forward another incoming flow of packets to satisfy QoS requirements.


 

%% file: Content/INC.tex
\section{A Primer of In-Network Computing}\label{INC}
\subsection{Beginnings}
\noindent In search of new solutions to improve network functionality, researchers sought mechanisms to inject customized programs -- encapsulated in “active capsules”-- into network nodes together with the classical data. In-packet encapsulated code is the main idea behind \textit{Active Networks}~\cite{ACTIVE} where network nodes actively process and perform calculations on the user data-in-transit.
As cloud-based services dominate the network landscape and ultra-low latency applications are emerging, the idea of active networks has been revived. Software-defined network (SDN) technology  provides excellent flexibility to implement network programming by separating control and data planes.
%
%
%
%
The SDN Controller interacts with the forwarding plane using the OpenFlow protocol~\cite{mckeown2008openflow}, whereas P4~\cite{bosshart2014p4} has been introduced to overcome some of the initial OpenFlow limitations such as the different traffic processing components and data path hardware designs that depend on switch vendors. A P4 program allows a packet header to be classified and an action to be executed on an incoming packet. Hence, we return to what was previously introduced as an active network and what is now known as in-network computing (INC).

In summary, in addition to approaches related to hardware acceleration, four major milestones have marked the development of INC: 
\begin{enumerate}
    \item Mid-1990s: Active networks introduced programmable functions in the network, paving the way to a paradigm shift in network architectures.
    \item 2004: SDN introduced the concept of control and data plane separation and open interfaces developed between them.
    \item 2010: OpenFlow API and network operating systems established large open interface adoption with advanced and scalable control and data plane techniques. 
    \item 2014: P4 was introduced as a step towards more flexible switches whose functions are specified and modified dynamically in operating networks.
\end{enumerate}

\subsection{An Illustrative Use-case}\label{sec:Usecases}
\begin{figure}
\centering
\includegraphics[scale=0.5]{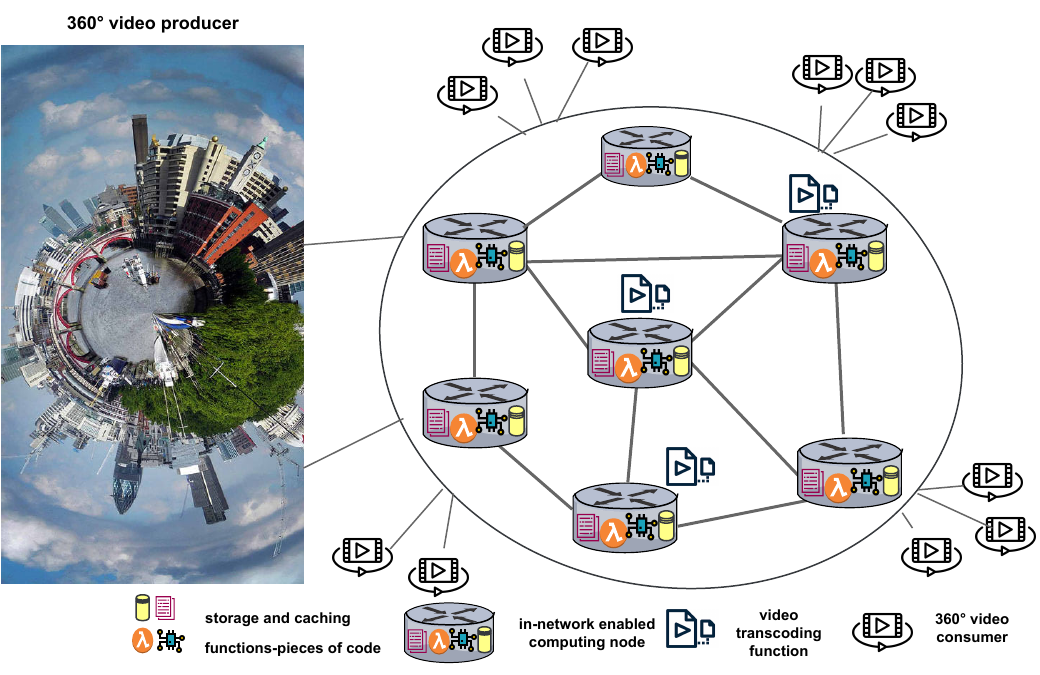}
\caption{360°video streaming using in-network computing}
\label{fig:INCusecase}
\end{figure}

\noindent In-network computation is used in a variety of fields to meet service requirements, such as packet aggregation, machine learning acceleration, network telemetry, and stream processing \cite{10.11453386367.3431302}.  
Figure \ref{fig:INCusecase} shows an illustrative use-case of the concept of in-network computing for delivering 360° video streaming applications. 360° video streaming has recently drawn increasing attention and publicity as an immersive technology. High bandwidth demand and intensive computational resource consumption make 360° video streaming on mobile devices challenging to achieve acceptable perceived quality. The video's visible area (known as the user's viewport)
 is displayed through a Head-Mounted Display (HMD) with a very high frame rate and high resolution. A pleasant 360° video streaming experience should prevent motion sickness, which can be met by highly responsive connectivity. The resolution of virtual reality (VR) immersive video in HMD needs to be extremely close to the amount of detail the human retina can perceive, which demands ultra-high bandwidth.
Video transcoding is a critical factor in media compress/decompress and up-sampling/down-sampling 360° videos. While centralized cloud-based computing and storage are not adequate for this type of latency-sensitive application. Edge-assisted streaming appears to be a promising solution to alleviate the poor cache hit ratio that increases the traffic volume.
To cope with this problem, \cite{8756899} has introduced edge transcoding implemented on the in-network computing resource. 
360° videos are transcoded to create multiple quality layers and tiles at the video stream provider's servers. 
The encoded video streams are then transmitted via HTTP to the Content Delivery Network (CDN). The solution proposed in \cite{8756899} allows content sharing by video frame tiling and load balancing through transcoding. In addition to potential latency reduction, a significant reduction (20 to 30\% ) in core network traffic volume can be achieved depending on the location of the transcoder.

Additional examples of in-network computing exist. For example, Vaucher et~\textit{al.} proposed an in-network data (de)compressor working at line-rate speed, allowing the offloading of the computations from end-nodes while still being suitable for time-critical applications \cite{10.11453386367.3431302}.  

\subsection{Key Definitions}
\noindent In this subsection, we classify the various INC definitions and provide a table describing how existing research describes INC.

\begin{table*}
\centering

\caption{INC Definition Classification}
\begin{tabular}{|c|c|c|c|c|c|c|c|} 
\hline
\multirow{3}{*}{\textbf{Research}} & \multicolumn{5}{c|}{\textbf{Definition of INC}}                                                                                                                                   & \multicolumn{2}{c|}{\textbf{Where to Process/Compute}}                         \\ 
\cline{2-8}
                                   & \multirow{2}{*}{\textbf{In-Network Processing}} & \multirow{2}{*}{\textbf{In-Network Computing}} & \multicolumn{3}{c|}{\textbf{In Switch x }}                                     & \multirow{2}{*}{\textbf{In-fabric}} & \multirow{2}{*}{\textbf{Out-of-fabric}}  \\ 
\cline{4-6}
                                   &                                                 &                                                & \textbf{Aggregation}     & \textbf{Caching}         & \textbf{Traffic Proc}    &                                     &                                          \\ 
\hline
\cite{ports2019should}                                &                                                 & \checkmark                       & \checkmark &                          &                          & \checkmark            & \checkmark                 \\ 
\hline
\cite{blocher2021holistic}                                 &                                                 & \checkmark                       & \checkmark & \checkmark & \checkmark & \checkmark            & \checkmark                 \\ 
\hline
\cite{mustard2019jumpgate}                                & \checkmark                        &                                                &                          &                          &                          & \checkmark            & \checkmark                 \\ 
\hline
\cite{mai2014netagg}                                & \checkmark                        &                                                & \checkmark &                          &                          & \checkmark            &                                          \\ 
\hline
\cite{anwer2015programming}                                  & \checkmark                        &                                                &                          &                          & \checkmark &                                     & \checkmark                 \\ 
\hline
\cite{10.1145/3152434.3152461}                                &                                                 & \checkmark                       & \checkmark &                          &                          & \checkmark            &                                          \\ 
\hline
\cite{tokusashi2018lake}                                &                                                 & \checkmark                       &                          &                          &                          & \checkmark            &                                          \\ 
\hline
\cite{Zilberman}                                &                                                 & \checkmark                       &                          &                          & \checkmark & \checkmark            &                                          \\ 
\hline
\cite{INC6G}                                 &                                                 & \checkmark                       & \checkmark & \checkmark & \checkmark &                                     & \checkmark                 \\ 
\hline
\cite{alim2016flick}                                  & \checkmark                        &                                                & \checkmark & \checkmark &                          & \checkmark            &                                          \\ 
\hline
\cite{zhang2014smartswitch}                                & \checkmark                        &                                                &                          & \checkmark & \checkmark & \checkmark            &                                          \\ 
\hline
\cite{InNet}                                & \checkmark                        &                                                &                          &                          & \checkmark &                                     & \checkmark                 \\ 
\hline
\cite{yang2019switchagg}                                &                                                 & \checkmark                       & \checkmark &                          &                          & \checkmark            &                                          \\ 
\hline
\cite{sapio2019scaling}                                & \checkmark                        &                                                & \checkmark &                          &                          & \checkmark            &                                          \\ 
\hline
\cite{jin2017netcache}                                 & \checkmark                        &                                                &                          & \checkmark &                          & \checkmark            &                                          \\ 
\hline
\cite{istvan2016consensus}                                 & \checkmark                        &                                                &                          &                          & \checkmark & \checkmark            &                                          \\
\hline
\end{tabular}
\label{tabe:INC Classification}
\end{table*}

The inconsistencies in the names used in the literature to describe INC may cause confusion. To lift the ambiguity of what INC refers to, we present an INC taxonomy based on the existing literature. As illustrated in Table~\ref{tabe:INC Classification}, we determine three main categories bearing resemblance to INC: 
(i) In-network processing, (ii) in-network computing/compute, (iii) in switch $X$, where $X$ is either caching, aggregation, data acceleration, replication protocols, or network sequencing.
We also distinguish two types of network devices for INC ~\cite{blocher2021holistic}: (i) Devices that execute the whole computing function, and (ii) devices that rely on a connected server for function execution. 

\subsubsection*{Taxonomy}
\textit{In-network processing} refers to processing data in network devices along its transmission on the path. Authors frequently use the term on path processing explicitly, like in the works of \cite{mai2014netagg} and \cite{InNet}, or implicitly like in \cite{mustard2019jumpgate} that defined in-network processing as to where data is processed by special-purpose devices as it propagates in the network. In-network processing is not limited to network devices, but it could be extended to middleboxes~\cite{anwer2015programming, alim2016flick}or Field Programmable Gate Arrays (FPGAs)~\cite{istvan2016consensus} which are considered by the authors as the best choice to process data at line rate. 
 
\textit{In-network computing/compute} refers to executing an operation or code on network devices or in virtual networks. Unlike in-network processing, in-network computing is used as an overarching concept and is often used in works that propose a complete architecture~\cite{10.11453359993.3366649} or a framework~\cite{INC6G}. Also, terms such as ``at line rate processing"~\cite{10.1145/3152434.3152461} and ``in fabric"~\cite{ports2019should} are used as synonyms to in-network computing on network devices. INC has been explicitly defined as the use of ASIC switches, FPGAs, or smartNICs for computing excluding network-linked accelerators in~\cite{NoaZilberman}.   
 
\textit{In switch $X$} refers to executing operations on the transmitted data like aggregation~\cite{InNet, 10.1145/3152434.3152461, sapio2019scaling, yang2019switchagg}, storing data on the network device such as caching~\cite{jin2017netcache}, or acceleration~\cite{tokusashi2018lake}. A switch is not a  conventional device used in the network, it could be virtual switches~\cite{INC6G} or newly developed switches~\cite{zhang2014smartswitch}.  

\subsubsection*{Deployment}
There are two basic types of deployment approaches for in-network computing devices, in-fabric and Out-of-fabric. 
In the \textit{in-fabric deployment}, computation occurs in network devices, such as programmable switches and Network Interface Cards (NICs). This approach offloads a set of computing operations from current computing nodes, e.g., servers, to the network elements such as the switches, routers, and smartNICs~\cite{10.1145/3152434.3152461}. Such an offloading accomplishes two benefits of in-network computing. First, capturing all traffic through a switch and NIC leads to a vision of the whole network. Second, in-fabric deployment does not add additional latency to the path that packets go through. However, as this method may have to support conventional routing and switching, there is a concern about the limitation of computation and memory resources. 

The second type, aka the \textit{out-of-fabric} deployment, is where in-network computing does not occur on network devices but separately from the fabric. The \textit{out-of-fabric} category is implemented using (i) a specialized hardware accelerator or (ii) exploiting Network Function Virtualization (NFV). This method does not give the same latency benefits compared to the in-fabric deployment. However, it leads to lower latency and higher throughput than the conventional approaches, where computation is performed on servers or devices outside the network. The immediate benefit is that this solution can be added without redesigning the entire data center network, i.e., the so-called incremental availability~\cite{Dan2019}.
\section{Technology Enablers} \label{INC_Enablers} 
\noindent In this subsection, we describe how the current advances brought by network softwarization such as Named Data Networking (NDN), SDN, and NFV technologies, enable the INC paradigm. We explain how NDN can be extended to support INC from a networking perspective. Then, we present the role of SDN and NFV in supporting INC from a virtualization perspective.

 \begin{figure}
    \centering
    \includegraphics[scale=0.41]{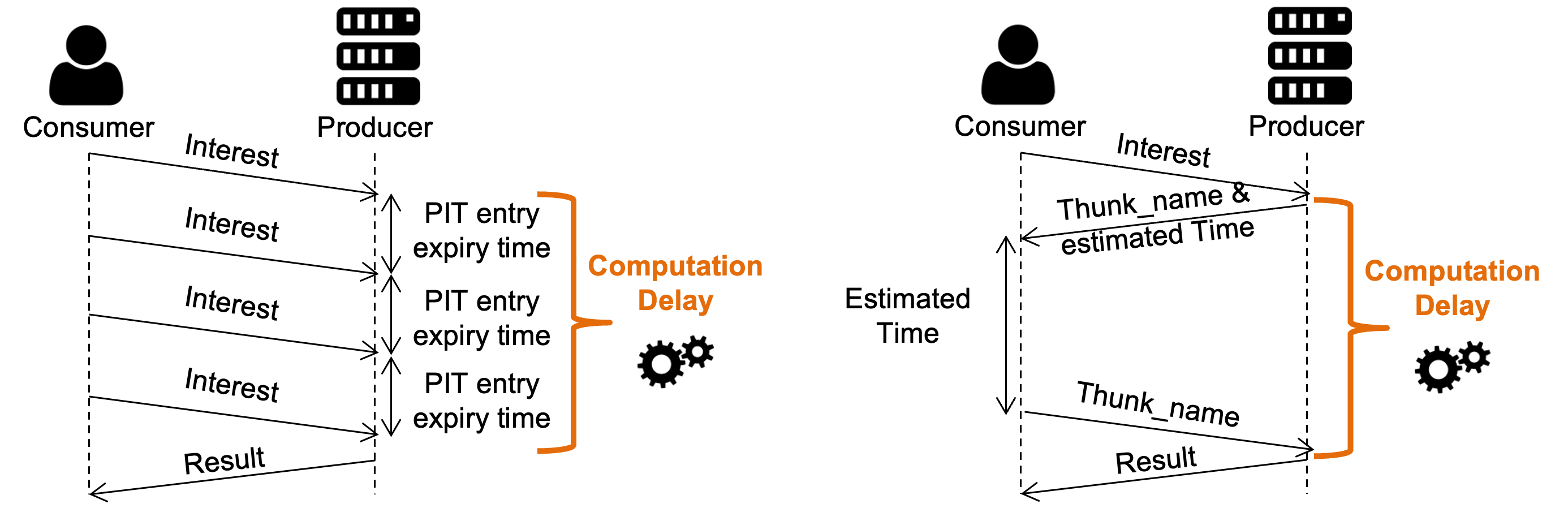}
    \caption{PIT entry expiry time in NDN: (left) the problem of PIT entry expiry time, (right) solution using thunk name}
    \label{fig:PITexpiry}
\end{figure}

\label{NDN}\subsection{NDN as a Networking Technology enabler}
\noindent NDN is a data-centric approach for networks that is based on location-independent names to find a requested content~\cite{NDNsurvey}.
As an alternative to the traditional TCP/IP-based Internet architecture, the NDN paradigm provides exciting features that can be extended to support in-network computing. For instance, NDN allows in-network content caching to respond to queries requesting existing content. This mechanism can be exploited to cache already calculated results to avoid redundant computations. In addition, NDN forwarding strategies can be enhanced to implement a location-independent execution and compute at the best network location~\cite{NFCC}.
There are two basic types of messages in NDN architecture that can be leveraged for in-network computing~\cite{NDNwireless}: \textit{Interests} representing requests for a particular computation to be executed and \textit{Named Data Objects} (NDOs) which can be a calculated result, parameters for the computation task, or a program code. The consumer sends an interest specifying the NDO to be fetched. The router then forwards this interest to the provider hosting the desired NDO. Each NDN forwarder involves three main components: (i) the Content Store (CS) that can cache NDOs (e.g., already calculated results), (ii) the Pending Interest Table (PIT) that is a storage of the interests requesting computation and the interface taken by these interests, and (iii) the Forwarding Information Base (FIB) which stores the name prefix of each execution result along with the following hops to reach the destination.

Notwithstanding the advantages offered by the NDN architecture, a few challenges associated with  NDN components for enabling in-network computing remain to be solved.\\
Firstly, network congestion is possible because of the large number of requests generated in network situations such as those of an IoT environment, and also because the network devices function as both forwarders and computers. Second, a PIT entry may no longer be valid after a specific time if a calculation takes longer time than the expiry time of the PIT entry. To overcome these challenges, \cite{COicn} proposed a framework that uses a computation-centric architecture over NDN with in-network load balancing of incoming computation requests and \textit{thunk names} to take into account the PIT entry expiry time. The thunk is a name used by consumers to retrieve execution results~\cite{RICE}. Figure~\ref{fig:PITexpiry} depicts the problem with the PIT entry expiry time and how it can be avoided using thunk names. In a traditional NDN architecture, a consumer sends a computing interest to the producer. Let us assume that the computation takes 10 ms and the PIT entry expiry time is 3 ms. The consumer first sends an interest that takes 3 ms, which expires later. Since the computation takes 10 ms, the consumer will be sending four interests before receiving the result. If the thunk name is used, the consumer sends the computation interest, and the producer responds with a thunk name and an estimated time. The consumer waits this estimated time and then fetches the result using the thunk name. based on these, the same work also demonstrates how in-network computing is implemented over NDN. The demonstration actors are Android smartphones (i.e., users), an access point providing wireless connectivity and performing in-network load balancing among the workers (e.g., Raspberry Pi). Workers are the devices responsible for in-network computing of users' requests. They retrieve the computation tasks as unikernels from the network. The use of unikernels for task execution over NDN is also proposed in \cite{10.1145/3125719.3125727}, where a framework based on serverless computing is proposed to extend NDN architecture in support of in-network execution of computation tasks (i.e., unikernels) via kernel stores. 
Based on a statistics table (i.e., a structure already existing in NDN), the kernel store decides which unikernels to store, which ones to remove, and which ones to execute.

 \label{NFV} \subsection{SDN/NFV as a Virtualization Technology enabler}

\noindent By incorporating INC into the Cloud-Edge-Mist continuum, off-loaded computing tasks are executed through an integrated infrastructure that optimizes network and computing resources. Each computational task can be decomposed into a set of subtasks. Each subtask can be realized as a virtualized function through containers or unikernels, deployed across the Cloud-Edge-Mist continuum. Virtualization is considered a major enabler of INC, as shown by Table~\ref{table:INC Framework}. According to previous related works, virtualization is often used to improve network overheads and accelerate network processing through virtual middleboxes. For instance, the authors of \cite{mai2014netagg} have proposed \textit{AGGBOX} to perform the aggregation along the path instead of the edge. \textit{AGGBOX} consists of a server connected by a high bandwidth link to network switches. Middleboxes are used to execute application-specific aggregation functions at each hop to reduce network traffic and solve the network bottleneck, thereby increasing application performance. 
Another use of virtualization is through virtual switches, e.g., in \cite{zhang2014smartswitch} the authors introduce a \textit{SmartSwitch} platform built on commodity servers connected with high-speed NICs. The proposed platform aims at simplifying the development of services that merge computation and storage into the network. The authors argue that \textit{SmartSwitch} can enable routers to perform complex processing on packets, or it could be integrated with a networking component to satisfy latency constraints. The authors first propose \textit{MemSwitch}, which they claim to be more flexible than the conventional switches by offering a dynamic way to control network redirection and load balancing. Secondly, their framework introduces a platform capable of moving storage from an endpoint to software-based middleboxes. Thirdly, \textit{MemSwitch} offers high-speed processing and filtering of data using memcach (i.e, generale distributed memory caching system). To summarize, the resulting \textit{MemSwitch} is a memcached-aware load balancing switch that decreases request latency, increases throughput, and allows in-network caching. 

On the other hand, in the era of cloud/edge computing and containerization, the virtualization of microservices has become very popular in the industry and academia because of its practical benefits. As microservices are self-contained services that perform a specific task, they are excellent candidates for in-network computing. Indeed, network devices such as switches have limited resources and may not be able to handle large monolithic services. Thus, it is better to break these down into microservices to make them smaller and more manageable~\cite{ENTICE}. 

In addition to the fundamental aspect of microservice architecture, namely scalability, it also addresses heterogeneity.  Virtualization allows the deployment of microservices on different devices of various capabilities in a heterogeneous infrastructure. Thus, virtualized microservices have been a key factor in improving cloud application performance~\cite{VirtualizationPerformance, 40}. In many studies, microservices are often implemented in containers because they are lightweight, efficient, and more deployable than traditional virtual machine (VM)-based implementation. In addition, container technology can be used to minimize the effects of virtualization on system resources and reduce its cost~\cite{46}. Implementing microservices as virtual functions taking benefits from INC would highly enhance both network and application performances.  

\begin{figure}[h]
\includegraphics[scale=0.47]{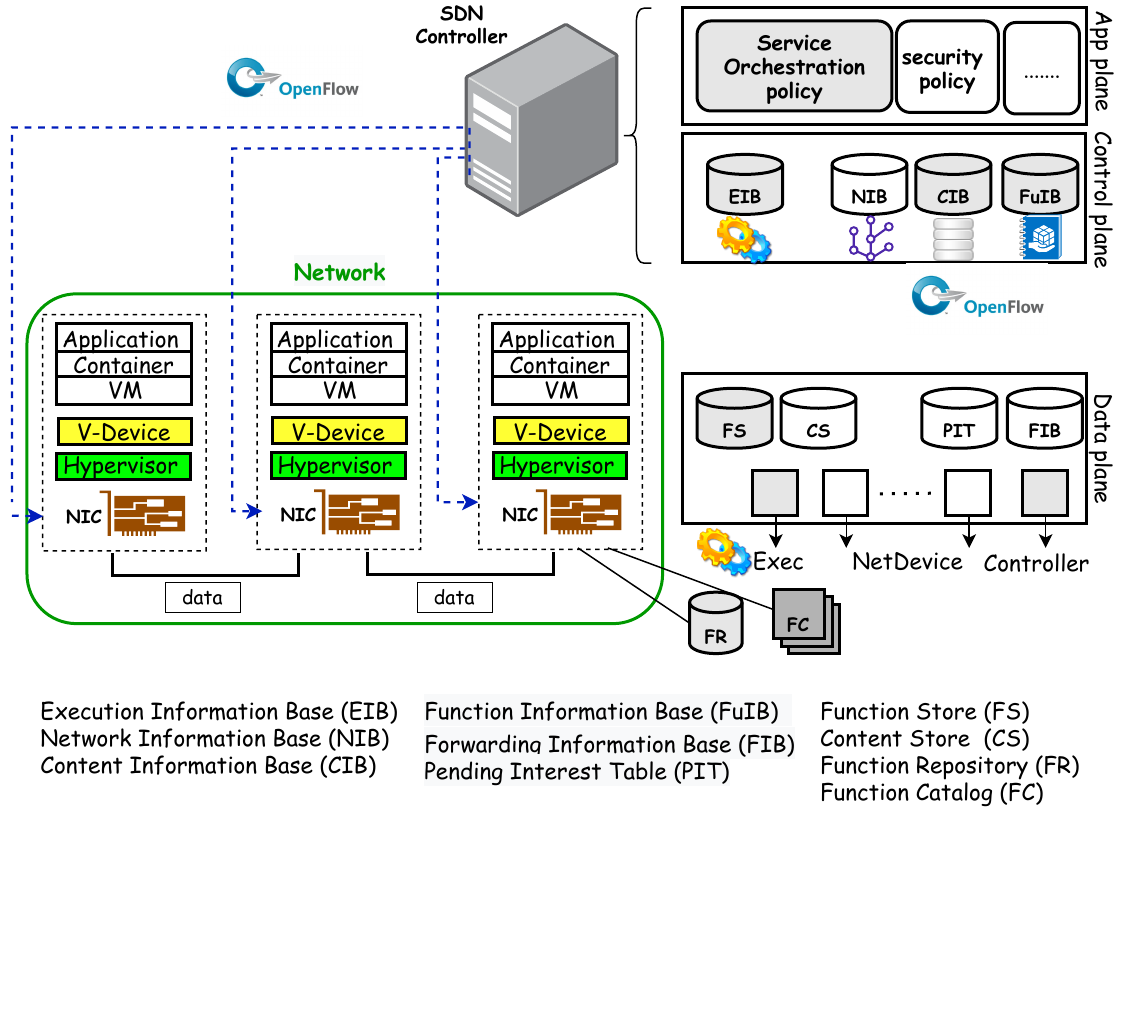}
\vspace*{-16mm}
\caption{In-network computing architecture using SDN, NFV and NDN (adapted from \cite{8856241} and \cite{10.11453445814.3446760})}
\label{fig:incArchitecture}
\end{figure}
The authors of \cite{8856241} proposed an SDN-based architecture of named computing services in edge infrastructures composed of three different planes: data, control, and application. The data plane contains the network nodes that make up the managed domain (i.e., ingress/egress nodes, backbone, intermediate, and access routers). The control plane includes the SDN controller managing its domain according to specific orchestration policies defined at the application plane. Moreover, a hierarchical control plane architecture can be envisaged for scalability purposes.\\
Following the architecture proposed in \cite{8856241} figure~\ref{fig:incArchitecture} illustrates the envisaged in-network computing architecture while making use of SDN, NFV, and NDN, where the gray components are the extensions to the conventional NDN proposed by~\cite{10.11453445814.3446760}. The three layers are described as follows.
\begin{itemize}
    \item \textit{Data plane:} In addition to NDN-specific data structures, each node keeps a Function Store (FS), specifically added to store the locally available functions that can be unikernels or containers. The SDN controller can specify a set of actions for the data plane, including (i) forwarding the interest to a given output port; (ii) contacting the controller; (iii) managing the computation locally. 
   \item \textit{Control plane:} For named service orchestration purposes, the controller needs to track the resource status of both the network and the nodes. Thus, the control plane implements two functionalities (i) network resources monitoring, and (ii) storage and computing resources monitoring.
    \item \textit{Application plane:} Adequate resource provisioning in such a system requires additional network applications compared to legacy NDN and SDN design, e.g., for security support and service orchestration. For instance, the authors in \cite{RICE, 8539024} have raised concerns that in-network computation creates new security challenges. In this work, first, the consumers should certify that the new contents are resulting from the computations. This can be ensured by specifying different security policies at the application plane and applied by the controller. Second, service orchestration decides where to execute a named computed request (executor node). The requested processing function can either exist locally at the FS of the selected executor node or be retrieved from a Function Repository (FR) hosted in a remote server known by the controller. Note that each FR is associated with a Function Catalog (FC)~\cite{ETSI}, which includes function descriptors (e.g., deployment template, required resources, etc.). This illustrates
    well how this plane works. 
\end{itemize}

Overall, in the INC context, the centralized control plane is primarily responsible for network optimization by calculating centrally optimized network policies and installing the corresponding flow tables on data plane devices to support processing and forwarding operations. A good example of such a plane is the context-aware pipeline designed for \textit{In-Switch $X$} targeting increase service satisfaction possibilities~\cite{li2021advancing}. The data plane first utilizes the context-aware caching and computing pipeline for the incoming packets to look for precomputed results that will significantly accelerate the service process and save communication and computation costs. In \cite{li2021advancing}, the SDN controller periodically collects important monitoring information (e.g., topology, traffic load, function lists, computation load, and content) to improve its policies. Moreover, the SDN controller installs extended fine-grained flow tables on data plane devices to support processing and forwarding operations of both interest and data packets.
Similarly, \cite{9350853} proposes an SDN-like collaborative control that periodically calculates the optimized task execution schedule for the entire network. Based on the centralized policies and status reports of the devices, the SDN-like collaborative control adjusts the forwarding strategy of the networking devices.\\
A classification of the aforementioned papers based on the used enablers (i.e., SDN or NDN), is given in Table~\ref{table:INC Framework}.

\begin{table*}
\caption{Current literature on INC frameworks}
\setlength\extrarowheight{5pt}
\label{table:INC Framework}
\begin{adjustbox}{width=\textwidth,center=\textwidth}
\resizebox{0.5\textwidth}{!}{
\begin{tabular}{ |c |c| c| c|}
\hline
\textbf{Papers} & \textbf{Enablers} & \textbf{Contribution}  & \textbf{How the enabler is used}\\
\hline
\cite{10.1145/3125719.3125727}& NDN & Named Function As A Service & Named function\\
\hline
\cite{RICE}& NDN & Remote Method Invocation in ICN & Named routing\\
\hline
\cite{INC6G} & SDN & In-Network Computing paradigm based on virtualized platform & SDN-like collaborative controller \\
\hline
\cite{li2021advancing}&SDN& A software-defined service-centric networking (SDSCN) framework & SDN Controller \\
\hline 
\cite{8856241}& SDN and NDN & \makecell{Managed Provisioning of Named Computing \\ Services in  Edge  Infrastructure} &\begin{tabular}[c]{@{}l@{}}  ~\begin{tabular}{@{\labelitemi\hspace{\dimexpr\labelsep+0.5\tabcolsep}}l@{}}Service Allocation decision mechanism based on SDN\\Request and delivery of Service  by Names\end{tabular}\end{tabular} \\

\hline
\cite{9350853} & SDN and NDN & A shared in network computing infrastructure &\begin{tabular}[c]{@{}l@{}}  ~\begin{tabular}{@{\labelitemi\hspace{\dimexpr\labelsep+0.5\tabcolsep}}l@{}}Name-based computing functions\\SDN-like collaborative schedule
controller\end{tabular}\end{tabular} \\
\hline
\cite{10.11453359993.3366649} &SDN and NDN& Software-Defined Service-Centric Networking (SDSCN) &\begin{tabular}[c]{@{}l@{}}  ~\begin{tabular}{@{\labelitemi\hspace{\dimexpr\labelsep+0.5\tabcolsep}}l@{}}Name-based forwarding functions\\Name-based function\\ Edge controller\\Cloud controller \end{tabular}\end{tabular} \\
\hline
\end{tabular}
} 
 \end{adjustbox}
\end{table*}

\section{Resource Provisioning in INC-enabled systems}\label{ResourceAllocation}
One of the main challenges in INC is where to place computing resources over the forwarding plane, this section aims to answer this question by describing the resource allocation mechanism from an INC functionalities perspective. We first survey existing work on computation entities placement and chaining i.e. VMs, containers, and functions. Next, task scheduling approaches and INC context peculiarities are summarized.


 
\subsection{Placement of computational entities}

\noindent The placement of computational entities has been widely studied in cloud environments with the goal of efficiently distributing services to data centers depending on their requirements in terms of latency, storage, and/or computation~\cite{cloud1}. However, it is largely admitted today that the techniques developed for cloud environments are not applicable to the heterogeneous and dynamic environments of edge computing. Likewise, the placement mechanisms proposed for edge and cloud computing may not be appropriate for in-network computing due to the nature of the underlying network devices, i.e., their significantly limited resources and computation capabilities~\cite{placeDelayINC, distributedINC}.
Application combining In-network and edge computing can be of great variety (e.g., AR/VR applications, online gaming, or autonomous robotics), so the orchestrator must be equipped with an adaptive placement strategy that allows effective decisions. The placement algorithms should take several challenges into consideration~\cite{placementSurvey}: (i) the heterogeneity of computational entities (e.g., virtual machines, containers, functions, and tasks), (ii) the variety of the equipment that can host these entities (e.g., servers, base stations, access points), (iii) the dynamicity of network conditions (e.g., mobility or immobility), and (iv) the different end-user requirements. In recent years, many efforts have been made to address these problems, from simple algorithms and heuristics to solving them with artificial intelligence techniques.

Placing a computing task in the network requires two essential components: (i) an executor node with available processing resources, and (ii) the transfer of the input data from the source to the selected executor node for processing. In highly distributed computing environments, where the application components consume computing, storage, and networking resources, the task and corresponding data placement algorithms substantially impact the performance. The placement strategies face two main challenges. First, (in general, a vast amount of) data exchange  via task executor nodes requires an adequate bandwidth provision. Second, task completion delays are intrinsically linked to end-to-end processing delays imposed by the location of task executor nodes.


Likewise, service function chaining (SFC) is recognized as an essential technique for connecting a sequence of SDN/NFV-based Service Functions (SFs). Achieving low-latency services entails reducing the SFC completion time through processing acceleration of softwarized SFs. Data plane reconfigurability and line-rate packet processing performance enabled by the Programmable Data Plane are vital to achieving that goal. 
Employing SFC in a programmable data plane can be done in two ways~\cite{FASE}. Firstly, a redundant service function approach that allows the easy processing of ordered SFs that make up an SFC is discussed in~\cite{P4SC}. In this approach, SF tables are redundantly embedded in the programmable data plane switch in order to satisfy the processing order of the SFC at the line rate. However, the limited resources of programmable data plane devices may not efficiently allow embedded service function redundancy. Second, the authors of ~\cite{AccSFC, RESFC} proposed the re-circulation approach when the order of SFC functions is different from the one of embedded SFs. A packet is recirculated from the egress to the ingress port. However, it is hard to guarantee the line-rate performance due to re-circulation. \cite{FASE} proposes a Flow-Aware Service function Embedding (\textit{FASE}) that balances these two drawbacks. It combines the redundant service function and re-circulation approaches to find the optimal service function embedding that minimizes the SFC completion time while the programmable data plane switch resources are efficiently utilized. 

From an architectural perspective, VNFs tend to be monolithic (e.g., load balancers, WAN optimizers), complicating their deployment in network devices. As INC embraces a heterogeneous infrastructure with differently sized hosts and networking devices, many of them with limited resources such as switches or smartNICs, the microservice architecture offers many attractive characteristics (e.g., heterogeneity and scalability) that accommodate service deployment in such environment. Applications can be decomposed into functionalities such as packet header parsing and packet classification~\cite{VNFMicro}. These small components can be easily integrated into network devices with limited capacity.
The authors of \cite{PIAFFE} have proposed a framework, called \textit{PIaFFE} (A Place-as-you-go In-network Framework for Flexible Embedding of VNFs), for VNF placement using in-network processing. In this work, the VNF logic is decomposed into embedded network functions \textit{eNFs} to identify the overlapped functions and efficiently place them in in-network programmable devices. \textit{eNFs} are compact and lightweight VNFs that use a high-level programming language and fit into a smartNIC.  
In addition, \cite{LightNF} proposes \textit{LightNF}, a system that allows network functions to be offloaded into programmable switches. The proposed system in \textit{LightNF} is equipped with an analyzer that examines the characteristics and options of offloading network functions and the resources consumed by each function. Then, based on this analysis, it builds an optimization framework that decides on the placement of SFCs.


To sum up all these, Table~\ref{placement} classifies the studies related to function placement according to four criteria. The first criterium is computation entities. Service function is a concept mainly used in service function chaining, where the service functions are connected to form a service that responds to a particular request. A task is an entity that focuses on one single action. Finally, embedded network functions are lightweight and small components of a VNF. The second criterium is the computing paradigm (In-network and/or edge computing) considered in each paper. The third criterium is the dependency of entities which refers to the connection (e.g., SFC) or the nonconnection (e.g., a single component) between the components. The last criterium pertains to the objective performance metrics considered in the placement of computational entities.


\begin{table*}
\setlength\extrarowheight{5pt}

\centering

\caption{Classification of works on placement}
\label{placement}
\begin{adjustbox}{width=\textwidth,center=\textwidth}
\resizebox{0.5\textwidth}{!}{
\begin{tabular}{|c|c|c|c|c|c|c|c|c|c|c|c|} 
\hline
\multirow{2}{*}{\textbf{Papers}} & \multirow{2}{*}{\textbf{Computation entities}} & \multicolumn{2}{c|}{\textbf{Computing paradigm}} & \multicolumn{2}{c|}{\textbf{Entities dependency}} & \multicolumn{6}{c|}{\textbf{Performance metrics}}                           \\ 
\cline{3-12}
                        &   & \textbf{INC} & \textbf{EC}  & \textbf{w/CXN} & \textbf{w/o CXN}           & \textbf{Latency} & \textbf{Throughput} & \textbf{Bandwidth} & \textbf{Energy} & \textbf{RU} & \textbf{Cost}  \\ 
\hline
\cite {LightNF}                      & Network Functions                     & \checkmark                  &                  & \checkmark           &                          & \checkmark     & \checkmark        & \checkmark       &        &                &       \\ 
\hline
\cite{distributedINC}                      & Tasks and computing resources         & \checkmark                  & \checkmark              & \checkmark           &                          & \checkmark     & \checkmark        & \checkmark       & \checkmark    &                & \checkmark   \\ 
\hline
\cite{FASE}                      & Service Functions                     & \checkmark                  &                  & \checkmark           &                          & \checkmark     &            &           &        &                &       \\ 
\hline
\cite{placeDelayINC}                      & Tasks                                 & \checkmark                  & \checkmark              &               & \checkmark                      & \checkmark     &            &           &        & \checkmark            &       \\ 
\hline
\cite{PIAFFE}                     & embedded Network Functions            & \checkmark                  &                  & \checkmark           &                          & \checkmark     & \checkmark        &           &        & \checkmark            &       \\
\hline
\multicolumn{12}{l}{INC: In-Network Computing, EC: Edge Computing, CXN: connexion, RU: Resource Usage.}
\end{tabular}
} 
 \end{adjustbox}
\end{table*}

 \subsection{Task Scheduling}

\noindent Task scheduling aims to distribute computing resources fairly and efficiently, thereby enhancing resource utilization. However, due to increasing workloads there are several challenges to be addressed. These challenges include load balancing, profit optimization, and software complexity~\cite{TSEC}. Considering INC, together with cloud and edge computing, adds extra complexity to the scheduler~\cite{TSVNF}. Hence, choosing which device to schedule computational tasks is crucial. The randomness of user request arrivals, physical device diversity, and their limited amount of resources should be taken into account when assigning tasks to devices. Additionally, each specific application has its requirements. For instance, it is necessary to consider task processing time, full utilization of resources, and bandwidth costs~\cite{TSVNF}.

\begin{figure}[!h]
\includegraphics[scale=0.7]{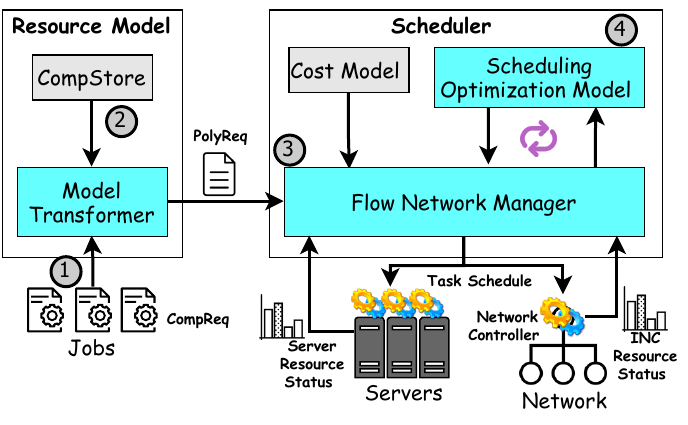}
\caption{HIRE Architecture (adapted from \cite{10.11453445814.3446760})}
\label{fig:hire}
\end{figure}

Although the resource management problem in traditional data centers (without INC) has been extensively studied, the problem is complex, and there is still no solution that fits all situations. The addition of the INC dimension to the resource pool not only exacerbates the existing challenges but also adds new ones~\cite{10.11453445814.3446760}. First, programmable network components (e.g., ASICs, FPGAs, and NPUs) are heterogeneous in terms of their programming models and interfaces compared to servers that support general Turing-complete computations. Thus, the same INC service exhibits different resource demands when deployed on different components. Second, INC services can request interchangeable resources with different performance properties. This results in extra complexity for scheduling appropriate resources for each INC job. Third, INC service comes with high locality constraints. Thus, the locality decision of a specific device impacts all others. Fourth, sharing INC resources may generate a nonlinear behavior, such as reusing partial INC resources by multiple services on the same network device. The authors of \cite{10.11453445814.3446760} address a few of these challenges by proposing \textit{HIRE} (Holistic INC-aware Resource managEr), a novel data center scheduler solution. \textit{HIRE} includes a scheduler that generates a flow network embedding all the scheduling constraints and objectives based on polymorphic resource requests (see Figure~\ref{fig:hire}). The scheduler considers the alternatives and non-linear resource sharing in the polymorphic requests. These requests represent the input of the \textit{HIRE} scheduler. The scheduler then generates a flow network based on a cost model that translates the current resource status, resource demands in polymorphic requests, and scheduling objectives into a flow network with costs on arcs.

Likewise, the authors in \cite{9350853} propose a shared in-network computing infrastructure named Intelligent Eco Networking (IEN). This infrastructure has two layers (i.e., the element and control layers). The control layer is responsible for scheduling the tasks executed in different nodes. However, the element layer is where compute tasks (i.e., name-based computing functions) are deployed. The authors also design an SDN-like collaborative schedule set in the slicing network layer along with name-based computing functions. The use of network slicing allows the proposed framework to coexist with the traditional TCP/IP network. The SDN-like collaborative schedule controller helps task scheduling and works as a global optimizer for the entire network. It runs dynamic programming algorithms for uncertain resources based on two strategies. Firstly, the multi-attribute decision-making strategy is scalable and adapts to the allocation of computing power while considering the static network and the dynamic computing resources. Secondly, information entropy is used for weight assignment to represent the dynamicity of network changes.

Finally, \cite{INC6G} proposes an in-network computing paradigm based on virtualization and software-defined technologies to realize unified scheduling and programming of computing tasks. The proposed paradigm allows the delegation of computing tasks to different nodes, including network devices, and rapid migration of tasks between network nodes. It also supports application-level service scheduling. The task scheduling of in-network computing is modeled as a multi-objective optimization problem. The optimization problem combines task scheduling and network forwarding to minimize the overall idle rate of resources, the overall energy consumption, and the overall data transmission overhead. 

Here again, Table~\ref{taskscheduling} summarizes the comparison between the works related to task scheduling. This comparison is based on the scheduling algorithm, the objective considered for scheduling, and the scheduling type (tasks and/or resource scheduling).

\begin{table*}
\setlength\extrarowheight{5pt}
\centering
\caption{Comparison of various task scheduling papers}
\label{taskscheduling}
\begin{adjustbox}{width=\textwidth,center=\textwidth}
\resizebox{0.5\textwidth}{!}{
\begin{tabular}{|c|l|l|c|c|} 
\hline
\multirow{2}{*}{\textbf{Papers}~} & \multicolumn{1}{c|}{\multirow{2}{*}{\textbf{\textbf{Scheduling algorithm}}}} & \multicolumn{1}{c|}{\multirow{2}{*}{\textbf{Objective}}}                                                                                                                                                                          & \multicolumn{2}{c|}{\textbf{\textbf{\textbf{\textbf{Scheduling}}}}}  \\ 
\cline{4-5}
                                  & \multicolumn{1}{c|}{}                                                        & \multicolumn{1}{c|}{}                                                                                                                                                                                                             & \textbf{Tasks} & \textbf{Resources}                                  \\ 
\hline
\cite{TSEC}                                & Resource Constrained Task Scheduling Profit Optimization algorithm (RCTSPO)  & Maximize profit                                                                                                                                                                                                                   & \checkmark            &                                                     \\ 
\hline
\cite{TSVNF}                               & Greedy Available Fit (GAF) task scheduling algorithm                         & Increase operational efficiency                                                                                                                                                                                                   & \checkmark              & \checkmark                                                   \\ 
\hline
\cite{TSVNF}                               & Heuristic resource management \textit{HealthEdge}                            & Minimize task processing time                                                                                                                                                                                                     & \checkmark              &                                                     \\ 
\hline
\cite{10.11453445814.3446760}                                & Flow-based scheduling using Min-Cost Max-Flow (MCMF) solver                  & Maximize job success rat                                                                                                                                                                                                          & \checkmark              & \checkmark                                                   \\ 
\hline
\cite{9350853}                                & Multi-attribute decision-making and information entropy based algorithm      & Adapt to dynamic changes                                                                                                                                                                                                          & \checkmark              &                                                     \\ 
\hline
\cite{INC6G}                                & Multi-objective evolutionary algorithm based on decomposition                & \begin{tabular}[c]{@{}l@{}}Minimize the overall:~\\\begin{tabular}{@{\labelitemi\hspace{\dimexpr\labelsep+0.5\tabcolsep}}l@{}}Idle rate of resources\\Energy consumption\\Data transmission overhead\end{tabular}\end{tabular} & \checkmark              &                                                     \\
\hline
\end{tabular}
} 
 \end{adjustbox}
\end{table*}

\section{Implementation Approaches}\label{Implementation}
\noindent SDN has brought to reality a novel paradigm in networking: programmability. Traditionally, the actions of programmable switches are driven by matching patterns in the tables, which can range from using only MAC addresses to more complex OpenFlow rules. 
\begin{figure}[!h]
\centering
\includegraphics[scale=0.63]{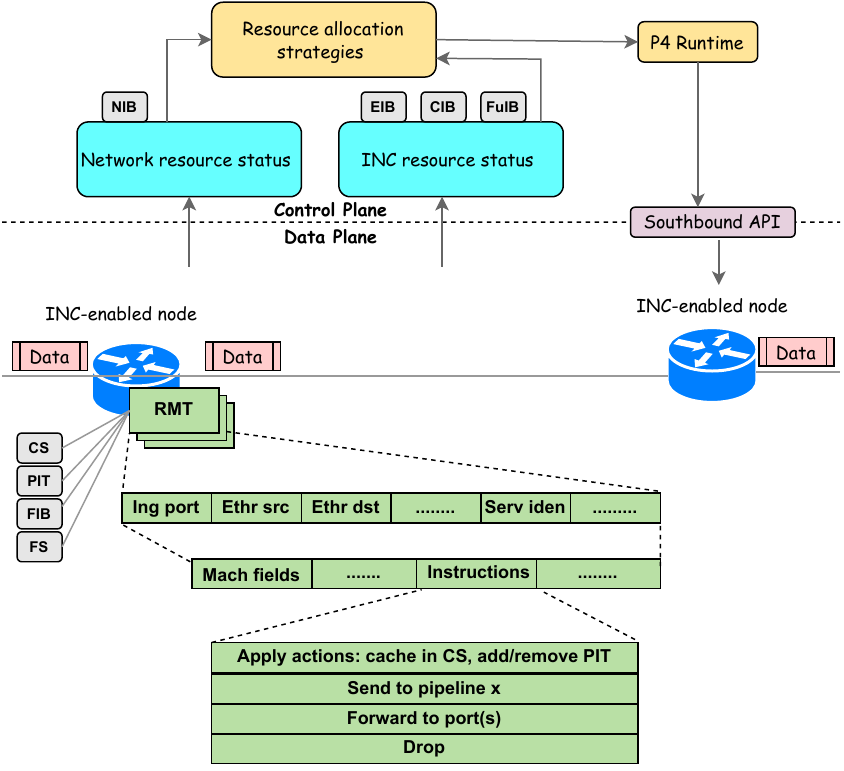}
\caption{Software defined networking based in-network computing model}
\label{fig:SDNenabler}
\end{figure}
However, this standard match-action paradigm does not apply to the INC context as emerging applications with diverse requirements demand customized matching criteria with corresponding device behavior. 
With the emergence of P4 \cite{bosshart2014p4}, the programmability of the network data plane is further extended to allow the network manager to express a novel matching pattern with a corresponding device behavior. As shown in Figure~\ref{fig:SDNenabler}, the incoming network traffic goes through a pipeline process that consists of Reconfigurable Match-Action Tables (RMTs)~\cite{bosshart2013forwarding}. RMTs map keys (e.g., packet headers) to actions (e.g., set egress port), while the control plane designs the optimal forwarding strategies (i.e., tables entries). Moreover, based on the abstract view of the topology, the control plane is capable of designing optimal \textit{In-Switch $X$} policies. Hence, the data plane first utilizes the caching and computing pipeline to look for pre-computed results in content storage that can significantly accelerate the service process and save the communication and computation costs~\cite{li2021advancing}. Hence, the maturity of SDN and P4 makes the vision of INC more practical~\cite{bosshart2014p4}.

\subsection{Named Function Networking (NFN)} \label{NFN}
\noindent NFN is an extension of Information-Centric Networking (ICN) by adding the means to support function definition and application to data while applying the same resolution-by-name process~\cite{sifalakis2014information}. It builds on $\lambda$-functions to execute services anywhere in a network device. However, as NFN is restricted to the basic $\lambda$-functions included in the Interest name, the number of supported services turns out to be limited. Expressing advanced processing or custom code in a network device is challenging when using only $\lambda$-functions. 
 
\begin{figure}
\centering
\includegraphics[scale=0.75]{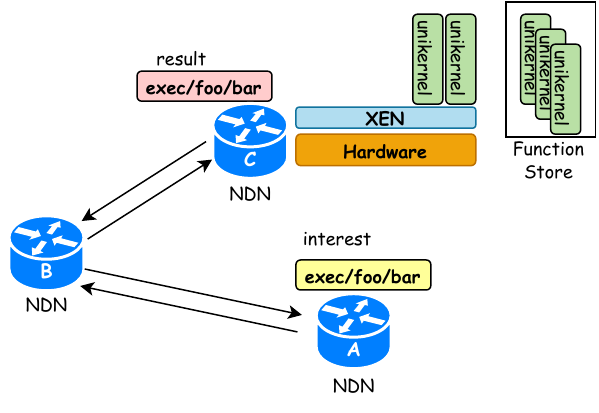}
\caption{NFN architecture}
\label{fig:NFNarchitecture}
\end{figure}

The most important work that brings the benefits of NFN to INC while avoiding the above shortcomings is the Named Function as a Service (\textit{NFaaS}), a framework proposed in \cite{10.1145/3125719.3125727}. NFaaS extends the NDN architecture using explicitly named unikernels to support in-network function execution. 
\textit{NFaaS} includes a Kernel Store component, which is responsible not only for storing functions but also for making decisions on which functions to run locally. In addition, it includes a routing protocol and several forwarding strategies to deploy and dynamically migrate functions within the network. 
The unikernel (i.e., function code) executes the requested action and wraps the response in a data packet that follows the transmission path to the requesting node. This assumes that the network node executing the request has enough computation capacity and storage capacity to cache the unikernel in its own storage. Otherwise, the interest request is simply forwarded to the next node. Figure~\ref{fig:NFNarchitecture} presents an overview of the system. An interest to execute the function \textit{/foo/bar} issued by node A is forwarded to node C, where the corresponding kernel is instantiated. The resulting data is sent back following the same path.
 
The authors of \cite{8264868} take advantage of NFN to stream and process data in the network leading to what they define as a  \textit{stream aware network}. The network is responsible for deciding when and where to process the stream, and the results are stored in the network nodes. Data streams are processed on the direct path between the source and the requester by optimising location based on mobile function codes (i.e., the functions are mobile and are executed in the best-located processing unit and not at the requester device). The proposed solution (i.e., the mobility of the function code) promises to reduce load compared to the traditional approach, as it distributes mainly mobile cloud codes, so the calculation can be distributed over network nodes.
Another work based on NFN was introduced in  \cite{scherb2017execution} to ensure a mechanism that will enable steering long computations running in the network over NFN. The proposed solution answers the limitation of NFN regarding the incapability of accessing or interacting with the computation once it is launched over the network, which does not stop until it delivers the results. The authors have addressed this problem by proposing \textit{Request 2 Compute} messages as an extension to NFN. Those messages enable users to access 
``process control blocks" and are used to request intermediate results to obtain the current state of a long-running application such as simulation. Table~\ref{INCdeploy} summarizes these studies  highlighting their contributions and design goals.

\subsection{In-Network Programmability}\label{Prog} 

\noindent In the SDN-based network computing model, the application logic fragment can be implemented through P4 with the aim to be deployed on a switch and its flow table. Here, the SDN controller plays a crucial role in deciding the optimal placement of processing and task scheduling \cite{tokusashi2019case}. Several research works propose optimal processing placement strategies by leveraging centralized control plane capabilities. For example, the In-Network Computing Architecture (\textit{INCA}) instantiates the requested function at appropriate places to meet the QoE constraint is described in \cite{10.11453359993.3366649}. In \textit{INCA}, the controller keeps track of the processing workload of the networking devices as well as the optimal routes in between them. Another work \cite{ruth2018towards} utilizes SDN and P4 in industrial control systems to offload delay-sensitive control tasks from the cloud to local network elements which can perform processing with bounded delay.  

Likewise, data aggregation and in-network caching are the most prominent efforts in the literature, leveraging SDN to implement INC services. The controller is primarily responsible for defining the aggregation tree and sending the corresponding processing rules to the switch in the form of table entries~\cite{10.1145/3152434.3152461}. For the incoming data packets, the switch parses the message and then aggregates the data according to the flow rules defined by the controller. Similarly, another framework is proposed in \cite{yang2019switchagg} to integrate the packet load analysis module and data aggregation processing engine into the switch, thus, improving the network throughput with reduced latency. Besides data aggregation, the controller controls the state consistency by maintaining an updated registration table on the switches in the data center network~\cite{liu2017incbricks}. As the popularity of each cached key-value pair changes over time, the central control is used to keep the current hot cached items on the switches. \cite{jin2017netcache}. In practice, network operators use programmable devices to modify the functionality of the data plane as needed while maintaining the ability to interpret, and process data flows at link speed. 
%
Use cases like data caching~\cite{jin2017netcache}, load balancing~\cite{grigoryan2019iload}, Domain Name System  server~\cite{woodruff2019p4dns}, and data aggregation~\cite{sapio2019scaling, yang2019switchagg} can be implemented within a network device, reducing the traffic in the network and consequently improving the overall performance.

Match-Action Tables allow network operators to use a set of pipeline stages containing a match table of width and depth arbitrary to match a particular field in an incoming packet on the pipeline. Programmable Data Planes (PDPs) enable network devices to perform complex operations on packets allowing the modification of how data is processed at the hardware level. Besides, Domain-Specific Languages created for PDP, such as Protocol-Oblivious Forwarding ~\cite{song2013protocol} and P4~\cite{bosshart2014p4}, allow programming the data plane behavior in a higher-level abstraction. Domain-specific language provides novel approaches in PDP \cite{da2017data}, helping to overcome even more  "ossification" of computer networking and allowing the development of a new generation of network services. Since P4 is the most prevalent enabling technology that is used to define data plane algorithms, the real power of P4 lies in a programmable match-action pipeline, describing tables, lookups, and actions in an abstract, straightforward manner with the freedom of defining any kind of protocol headers.

On the other hand, P4 is a domain-specific language that does not support complex operations such as checksum or hash computation units, random number generators, packet and byte counters, meters, registers, and many others. To make such extern functionalities usable, P4 introduces the so-called externs. These externs are  the external objects describing the interfaces that such objects expose to the data plane. On a separate note, the implementation of complex operations has been investigated as extensions to P4. In~\cite{da2018extern}, a case study is performed on integrating the Robust Header Compression (ROHC) scheme, which leads to an implementation of a new native primitive. Another work, \cite{scholz2019cryptographic}, proposes the extension of P4 by cryptographic hash functions that are required to build secure applications and protocols. The authors of \cite{scholz2019cryptographic} have proposed an extension of P4 Portable Switch Architecture and implemented a PoC for three platforms: CPU, Network Processing Unit (NPU), and FPGA-based P4 targets. In addition, the asynchronous execution of externs has been investigated in \cite{laki2020price, horpacsi2019asynchronous}, showing that packets may be processed by the pipeline during the execution of externs.

\begin{figure}[h]
\includegraphics[scale=0.7]{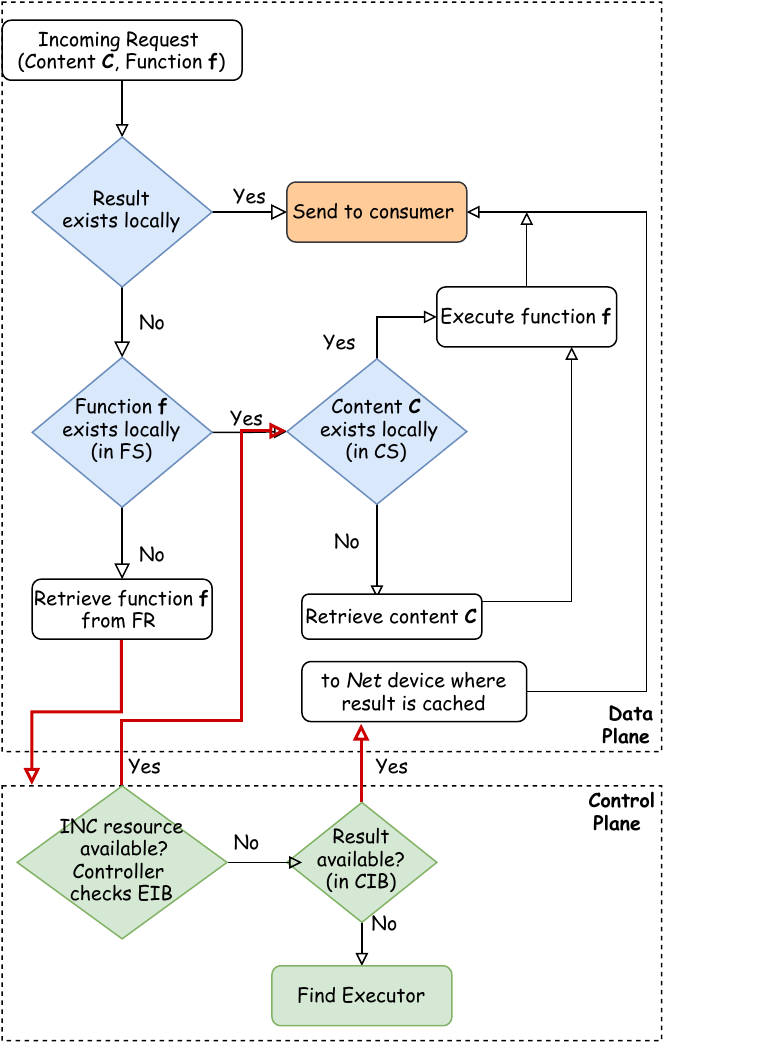}
\caption{NFN and SDN in action}
\label{fig:NFN}
\end{figure}
\subsection{Summary}\label{summaryEnablers}
\noindent To illustrate the different actions of the INC enablers, we draw a flow chart in Figure~\ref{fig:NFN} summarizing the interest request processing steps from its generation by a consumer to the reception of the final result. 
Suppose that an interest request composed of content $C$ and function $F$ arrives at the first node (node $N$), then the node checks if the results are stored in its content store CS from previously executed computations. If the requested result exists, it is sent back to the consumer; otherwise, the node looks for function $F$ in its function store FS and content $C$ in CS. 
Failure to match function $F$ implies retrieving it from the function store FS. Note that FS is hosted in a remote server known by the controller, as described in Subsection~\ref{INC_Enablers}. 
Likewise, the fail-matching of content $C$ follows the classical NDN procedure for finding content in the network and moving it to node $N$ where the execution of the interest request (content $C$, function $F$) occurs. 
If both $F$ and $C$ exist at node $N$, then the execution can proceed if and only if there is enough available CPU and caching capacity. A request message is sent to the controller to obtain INC resource status from the execution information base (EIB). If there is capacity, the execution proceeds; otherwise, the controller scans the content information base (CIB) to verify if the required result exists in other nodes. Otherwise, the controller will launch the action of finding an executor node through the resource allocation strategy module (see Figure~\ref{fig:SDNenabler}).
\begin{table*}
\centering
\caption{INC deployment frameworks\label{INCdeploy}}
\begin{tabular}{ |c |c| c| c|}
\hline
\textbf{Papers} & \textbf{Architecture} & \textbf{Design goals} & \textbf{Contribution} \\
\hline
\cite{10.1145/3125719.3125727} & NFN & In-Network Function execution& \begin{tabular}[c]{@{}l@{}}In-Network Function:  :~\\\begin{tabular}{@{\labelitemi\hspace{\dimexpr\labelsep+0.5\tabcolsep}}l@{}}Storing\\Dynamic deploying\\Migration\end{tabular}\end{tabular}\\
\hline
\cite{da2020pipelining} & NFN & In-Network Stream processing (Stream aware network) &  \begin{tabular}[c]{@{}l@{}}In-Network Stream:  :~\\\begin{tabular}{@{\labelitemi\hspace{\dimexpr\labelsep+0.5\tabcolsep}}l@{}}Storing\\Processing\\\end{tabular}\end{tabular} \\
\hline
\cite{scherb2017execution} &NFN& Steering long computation over the network& \makecell{Extension of NFN by adding\\ \textit{Request2Compute} messages}  \\
\hline
\cite{da2018extern} &P4& Supporting header compression and decompression& \makecell{Add new external primitives to P4\\ to support header compression and decompression (ROHC)} \\
\hline
\cite{kim2020unleashing} &P4& \makecell{ Exploiting INC \\ to accelerate Scientific application workload }& NSinC (Network-accelerated ScIeNtific Computing) \\
\hline
\cite{scholz2019cryptographic}&P4& \makecell{ Enabling authentication and resiliency\\ in P4 targets} & Extending P4 with external cryptographic hash function \\
\hline
\end{tabular}
\end{table*}

%% file: Content/challenges_opportunities_C4.tex
\section{Challenges, Opportunities, and Future Research Directions}\label{Challenges}
\noindent The in-network computing concept is changing the way we perceive the next-generation networking infrastructure. We can now reevaluate where and when data is processed and cached, and how and what is communicated. This paradigm shift has its own challenges, from hardware implementation and architectural design to regulatory and commercial aspects; and its opportunities from autonomy to scalability and to applying AI within the network. In the following, we will review each challenge as well as the opportunities for future research. 

\subsection{Technical Challenges for INC}
\vspace{-3pt}
\subsubsection{Memory} \noindent Recently, the advances in programmable data plane devices open up many opportunities to offload operations or functionalities to the switches. However, the limited memory constraint of the switches remains a major challenge that many applications make struggles to deal with it. Besides, the applications mentioned in Section~\ref{sec:Usecases} face a clear trade-off between performance and memory usage. For instance, in the in-network cache scenario, the cache hit rate and the accuracy of sketches are directly determined by the memory size of the networking equipment~\cite{jin2017netcache, liu2016one}. Similarly, the memory exhaustion of the device directly leads to a slow down of the performance of applications such as load balancing~\cite{miao2017silkroad} and monitoring~\cite{narayana2017language}. Even the most basic function of switches, such as packet forwarding, can suffer from limited memory space. To tackle the memory constraint, one solution is the design of a switching ASIC with internal custom logic and wires for the external Dynamic Random Access Memory (DRAM) access. However, this proves to be costly due to the design of numerous parallel DRAM modules combined with a DRAM controller. Therefore, the on-switch external DRAM is not widely used by vendors in today’s single-chip switches. Another possible solution is to utilize the unused resources and DRAM of the data center networks. However, there are some technical challenges in terms of performance, fault tolerance, and load balancing, though the programmable switch memory can be extended in a cost-effective manner. For example, the work in \cite{kim2020unleashing} particularly explores the possibility of re-purposing the affordable DRAM installed on data center servers to expand the memory available on a switch’s data plane. Instead of building another switching ASIC with a custom external-DRAM-access capability, they simply reuse an existing programmable switching ASIC built only with on-chip memory. The proposed approach designs three remote memory primitives: remote packet buffer, remote lookup table, and remote state store, leveraging Remote Direct Memory Access (RDMA) operations. The experimental result shows that the proposed system achieves high performance in terms of bandwidth, causing a little latency cost and absolutely 0\%  CPU overhead.

\subsubsection{Power consumption} Power consumption of programmable devices is one of the key performance indicators that directly reflect the packet processing efficiency of the chip,  measured in million packets per second per Watt (Mpps/Watt)~\cite{pongracz2013cheap}.
Unlike the traditional chips with a lower number of transistors and switching frequency, today’s chips are of smaller volume and embed a large number of transistors that allow them to leave a higher processing capacity. Hence, it leads to more power dissipation which directly reflects the computing time and generates additional cooling costs. To overcome this power dissipation, the work in \cite{yang2019switchagg} redesigned the routing switches. These switches operate in two modes: (i) sleep mode that reduces the leakage power, and (ii) low-power mode that reduces the dynamic power.

\subsection{Adding the control dimension to INC}
\noindent Even though a general theory of goal-oriented communication was put forward a decade ago~\cite{10.1145/2160158.2160161}, this has not been fully achieved yet. However, transitioning to goal-oriented communication is becoming more urgent since end-users have become \textbf{producers} and \textbf{consumers} of contents as their devices are now embedded with various sensors, which help in creating and collecting various types of data from different domains such as energy, agriculture, healthcare, transport, security, and smart homes, among others.  With the exponential growth of diverse IoT devices, as well as various applications (e.g., intelligent manufacturing, smart agriculture), the expectations for the performance, reliability, and security of networks are greater than ever. For example, an industrial control system requires a strict target of a 2 ms network delay and 1 ms jitter. While achieving a controllable and observable system with such target performance metrics has been studied over more than half a century, and various criteria have been developed, difficulties arise when they are straightforwardly applied to a networked system consisting of a large number of subsystems. To be more specific, it has been recognized that for large-scale networked systems, these criteria are usually computationally prohibitive.

To this end, INC has the potential to alleviate the complexities of networked control systems by recognizing that communication is not an end in itself but rather a means to achieve some goals of the communicating parties. 
%
%

In this aspect, INC can enable 4C, which is the extension of 3C with the addition of physical controllers. The act of control, is the end result of a collection of functions that operate at the management plane, and includes the computing, caching, and communication (3C) functions and their coordination to ensure an efficient operation.  These controllers could be deployed as microservices and placed optimally within the infrastructure alongside the 3C functions. That's where INC can come into play. For instance, a lifecycle management function can be deployed close to a compute function to optimize the control and management of that function on the run, without the delay induced when having a remote management entity.

\label{INC_business}
 \noindent 

\begin{table*}[ht]
\centering
\captionsetup{format=hang}
\caption{Regulatory, Privacy, Safety, Commercial, Legal and Ethical Issues Classification}
\label{table:Challenges_RSE}
\begin{adjustbox}{width=\textwidth,center=\textwidth}
\resizebox{0.5\textwidth}{!}{
\begin{tabular}{|c|c|c|c|c|c|c|c|c|c|} 
\hline
\multirow{2}{*}{\textbf{Papers}} & \multicolumn{2}{c|}{\textbf{Regulatory }}             & 
\multicolumn{2}{c|}{\textbf{Privacy }}             &
\multicolumn{1}{c|}{\textbf{Safety }}                                             & \multicolumn{3}{c|}{\textbf{Legal} \& \textbf{Ethical Issues} }  &  \multicolumn{1}{c|}  {\textbf{Commercial} }   \\ 
\cline{2-9}
                                   & \textbf{Distributed Learning}      & \textbf{Knowledge Transfer}        & \textbf{Privacy Protection}        & \textbf{Differential Privacy}        & \textbf{Secure Computation}      & \textbf{Data Ethics}               & \textbf{Applied Ethics}            & \textbf{Business Ethics}      &    \\ 
\hline
\cite{gao2021sedml}                                 & \checkmark & \checkmark & \checkmark & \checkmark & \checkmark &                           &                           &                           & \\ 
\hline
\cite{9629226}                                  &                           & \checkmark &                           &                           & \checkmark &                           &                           &                          &  \\ 
\hline
\cite{anahideh2021choice}                                  &                           & \checkmark &                           &           \checkmark                &  &                           &                           &                           & \\ 
\hline
\cite{8662743}                                &                           &                           &                           &                           &                           & \checkmark & \checkmark &       &                     \\ 
\hline
\cite{9293092}                                &  \checkmark                 &                           &                           &                           &     \checkmark                      &  &     &  &                          \\ 
\hline
\cite{8768490}                                 &                           &                           & \checkmark & \checkmark & \checkmark &                           &                           &            &                \\ 
\hline
\cite{8789542}                                &                           & \checkmark & \checkmark &                       \checkmark    &  &                           &                           &                      &      \\
\hline
\cite{zhan2020incentive}&                           &  &   &                          &  &                           &                           &                      &     \checkmark  \\
\hline
\cite{tang2020communication}                                &                           &   &   &                          &  &                           &                           &                      & \checkmark  \\
\hline
\end{tabular}
} 
 \end{adjustbox}
\end{table*}

\begin{table*}
    \centering
    \caption{Regulatory, Privacy, Safety,Legal and Ethical Issues Table of Acronyms}
    \begin{tabular}{|c|c|}
    \hline
    \textbf{Acronym} & \textbf{Term} \\
    \hline
        ISP & Internet Service Provider \\
        \hline
        INC & In-Network Computing \\
        \hline
        NGNI & Next Generation Network Initiative \\
        \hline
        CEP & Complex Event Processing \\
        \hline
        AI & Artificial Intelligence \\
        \hline
        DL & Deep Learning \\
        \hline
        SEDML & Securely and Efficiently Harnessing Distributed Knowledge in Machine Learning \\
        \hline
        SLA & Service-Level Agreement \\
        \hline
    \end{tabular}
    \label{tab:Acronyms_RSE}
\end{table*}
 \subsection{Regulatory, Privacy, Safety, Legal and Ethical Issues}
 
 \subsubsection{Regulatory}

INC has the potential to improve the overall system performance in Next Generation Networks Infrastructures (NGNI) by reducing processing time and increasing bandwidth efficiency. INC is steadily changing ``forward-only" routers into server-like devices capable of storing and/or processing incoming data. In other words, as storage and computing grow more affordable, in-network devices will be able to function as standalone servers more frequently. Hence, as computation and caching become new functionalities for ISPs, the network neutrality regulations will need to be reevaluated.  
Note that the regulations will affect the rules on which computing tasks should be transferred to INC devices and which should be retained. This can be handled by Complex Event Processing (CEP) ~\cite{9293092} by transforming the application-specific function into some simple ``match-action" processes for capturing and analyzing data streams to figure out what went wrong (i.e., complex events). Another example is a system created in \cite{9629226} that pools mid-path storage and connects resources to transport data across the end-to-end path. Along the way from the original server to the client, data can transfer from one hop to the next with this new system. It contains a set of mechanisms that ensures: (i) zero packet loss in intermediate network routers, (ii) network stability, and (iii) higher link utilization.
 
 \subsubsection{Privacy}
In cloud computing, the information from all originating nodes is unnecessarily provided to the final destination, which may also create a privacy concern. In this aspect, INC presents a practical trade-off by achieving a privacy-preserving delivery ratio that is near to the ideal. The permission of distributed learning can be provided by subnetworks. Because there are no protocol boundaries between devices when they are all part of the same broadcast domain, imposing security policies is difficult. By breaking up a single large broadcast domain into multiple smaller broadcast domains, a boundary between devices is created.
 
 \subsubsection{Legal and Ethics}
INC will be essential in the support of immersive experience services. Immersive services require even more personal data such as physical attributes of the person to be shared in order to provide a customized service.  Legal and ethical conundrum exits on how this information will be anonymized but it can still be used for customized services.  Although the technical issue of machines for resolving ethical quandaries in the real world is clarified, a society-wide dispute may be necessary to consider and accept the protocols and rules for such services~\cite{8662743}. 
There are also detailed questions to be resolved, such as the appropriate legal framework when it may be necessary to reveal some of the data or its analysis. Examples include accident data for cars, medical malpractice, and criminal conspiracy.

\subsubsection{Safety and Trust}
An INC node must ensure that the service requests are generated from genuine end-user devices. At the same time, there must be a mechanism to measure the trust of a computing, caching, or control service as well as what are the primary attributes that define the trust of the service. Mainly, a two-way trust relationship is desirable to develop a trusted interaction between INC nodes and end-user devices. An opinion-based model is helpful to choose a computing, caching, or control service. However, the reliability will become an essential factor to be considered. 

The integration of blockchain with the INC has the potential to enable end-to-end service delivery across the whole ecosystem. Without relying on an intermediary blockchain provides a distributed, unchangeable, single source of truth that everyone agrees with. When INC meets blockchain, two challenges of INC evolve with the advancement of NGNI because of geographically distributed edge nodes and those having limited capabilities. These two challenges that need to be handled are: (i) the assurance of multi-domain resource integration's trustworthiness, and (ii) the overall schedule of heterogeneous and geographically distributed edge and INC resources. For the first challenge, to assure reliability and traceability, blockchain has to be integrated with edge computing and INC. 

The use of INC and blockchain technology can effectively remove the requirement for trust in regulatory compliance. Removing the requirement for trust in regulatory compliance has a challenging technical-legal problem on top of a strong reliance on trust. It can be done by providing access to immutable data to both quality control authorities and enterprises. This will allow control authorities and enterprises to participate more fully in the process.

\subsubsection{Commercial}\label{INC_business}
INC is an inherently heterogeneous and distributed system and it is possible that different components are governed by independent and competitive autonomous entities.  When in-network computing is performed the cost/benefit should be shared by the related parties.  Although incentive designs and cooperative games are used extensively in the literature for efficient distributed resource allocation in the networks \cite{zhan2020incentive, tang2020communication}, INC introduces additional challenges to be resolved.  These challenges mainly arise due to the scale of operations in INC and it includes efficient and scalable accounting, authentication of operations to be completed, sharing of the incentives (which can be non-monetary).  

Table~\ref{table:Challenges_RSE}, classifies the above discussed papers according to the keywords which they are all mentioned directly or indirectly in papers. Finally, Table~\ref{tab:Acronyms_RSE} gives the acronyms used this section.

%% file: Content/Conclusion.tex
\section{Conclusion }
\noindent In this paper, we have shed light on the main distributed computing landscape in cloud-edge-fog-mist continuum. In particular, we have first presented a comprehensive overview of the fundamentals of In-Network Computing paradigm, including a taxonomy, technology enablers and implementation techniques. We have laid out a comprehensive summary of existing research on Computing, Caching and Communication (3C) integration and highlighted the role of In-Network Computing in shaping next-generation networks infrastructures (NGNI). 	
Based on these, we have scrutinized the necessity of enabling In-Network Computing with 3C integration to fulfill the stringent requirements of emerging applications. Specifically, through an extensive analysis of several use-cases, we have examined the synergy between INC and 3C integration to highlight the crucial role of INC in NGNI. Then, we have articulated the challenges of leveraging In-Network Computing in NGNI from different perspectives, including hardware implementation, architectural design, regulatory and business aspects.
Provided the insights concluded from this thorough tutorial, we conclude with several opportunities and future research directions to achieve the stringent requirements of futuristic applications and successful deployment of NGNI.  

%% file: old_stateofart.bib
@ARTICLE{8060515,
  author={Wang, Chenmeng and He, Ying and Yu, F. Richard and Chen, Qianbin and Tang, Lun},
  journal={IEEE Communications Surveys   Tutorials}, 
  title={Integration of Networking, Caching, and Computing in Wireless Systems: A Survey, Some Research Issues, and Challenges}, 
  year={2018},
  volume={20},
  number={1},
  pages={7-38}
}

@article{10.1145/2160158.2160161,
author = {Goldreich, Oded and Juba, Brendan and Sudan, Madhu},
title = {A Theory of Goal-Oriented Communication},
year = {2012},
issue_date = {April 2012},
publisher = {Association for Computing Machinery},
address = {New York, NY, USA},
volume = {59},
number = {2},
issn = {0004-5411},
url = {https://doi.org/10.1145/2160158.2160161},
doi = {10.1145/2160158.2160161},
journal = {J. ACM},
month = {may},
articleno = {8},
numpages = {65}
}

@electronic{ETSI,
  title         = "ETSI ENI (Experiential Network Intelligence) ISG (Industry
Specification Group)",
  url           = "https://www.etsi.org/technologies-clusters/technologies/experientialnetworked-
intelligence",
  year          = "2019",
}

@article{gao2021sedml,
  title={SEDML: Securely and Efficiently Harnessing Distributed Knowledge in Machine Learning},
  author={Gao, Yansong and Li, Qun and Zheng, Yifeng and Wang, Guohong and Wei, Jiannan and Su, Mang},
  journal={arXiv preprint arXiv:2110.13499},
  year={2021}
}

@article{anahideh2021choice,
  title={On the Choice of Fairness: Finding Representative Fairness Metrics for a Given Context},
  author={Anahideh, Hadis and Nezami, Nazanin and Asudeh, Abolfazl},
  journal={arXiv preprint arXiv:2109.05697},
  year={2021}
}

@ARTICLE{8789542,
  author={Zhaofeng, Ma and Xiaochang, Wang and Jain, Deepak Kumar and Khan, Haneef and Hongmin, Gao and Zhen, Wang},
  journal={IEEE Transactions on Industrial Informatics}, 
  title={A Blockchain-Based Trusted Data Management Scheme in Edge Computing}, 
  year={2020},
  volume={16},
  number={3},
  pages={2013-2021},
  doi={10.1109/TII.2019.2933482}}

@INPROCEEDINGS{8768490,
  author={Khalifa, Duoaa and Madjid, Nadya Abdel and Svetinovic, Davor},
  booktitle={2019 Sixth International Conference on Software Defined Systems (SDS)}, 
  title={Trust Requirements in Blockchain Systems: A Preliminary Study}, 
  year={2019},
  volume={},
  number={},
  pages={310-313},
  doi={10.1109/SDS.2019.8768490}}

@ARTICLE{9629226,
  author={Rene, Sergi and Ascigil, Onur and Psaras, Ioannis and Pavlou, George},
  journal={IEEE/ACM Transactions on Networking}, 
  title={A Congestion Control Framework Based on In-Network Resource Pooling}, 
  year={2021},
  volume={},
  number={},
  pages={1-15},
  doi={10.1109/TNET.2021.3128384}}

@ARTICLE{8662743,  author={Winfield, Alan F. and Michael, Katina and Pitt, Jeremy and Evers, Vanessa},  journal={Proceedings of the IEEE},   title={Machine Ethics: The Design and Governance of Ethical AI and Autonomous Systems [Scanning the Issue]},   year={2019},  volume={107},  number={3},  pages={509-517},  doi={10.1109/JPROC.2019.2900622}}

@article{10.1145/3406011,
author = {Schleier-Smith, Johann and Sreekanti, Vikram and Khandelwal, Anurag and Carreira, Joao and Yadwadkar, Neeraja J. and Popa, Raluca Ada and Gonzalez, Joseph E. and Stoica, Ion and Patterson, David A.},
title = {What Serverless Computing is and Should Become: The next Phase of Cloud Computing},
year = {2021},
issue_date = {May 2021},
publisher = {Association for Computing Machinery},
address = {New York, NY, USA},
volume = {64},
number = {5},
issn = {0001-0782},
url = {https://doi.org/10.1145/3406011},
doi = {10.1145/3406011},
journal = {Commun. ACM},
month = {apr},
pages = {76–84},
numpages = {9}
}

@ARTICLE{9403911,  author={Cecchinato, Davide and Erseghe, Tomaso and Rossi, Michele},  journal={IEEE Transactions on Network Science and Engineering},   title={Elastic and Predictive Allocation of Computing Tasks in Energy Harvesting IoT Edge Networks},   year={2021},  volume={8},  number={2},  pages={1772-1788},  doi={10.1109/TNSE.2021.3072968}}

@article{sliper2020energy,
  title={Energy-driven computing},
  author={Sliper, Sivert T and Cetinkaya, Oktay and Weddell, Alex S and Al-Hashimi, Bashir and Merrett, Geoff V},
  journal={Philosophical Transactions of the Royal Society A},
  volume={378},
  number={2164},
  pages={20190158},
  year={2020},
  publisher={The Royal Society Publishing}
}

@article{FIROUZI2021101840,
title = {The convergence and interplay of edge, fog, and cloud in the AI-driven Internet of Things (IoT)},
journal = {Information Systems},
pages = {101840},
year = {2021},
issn = {0306-4379},
doi = {10.1016/j.is.2021.101840},
author = {Farshad Firouzi and Bahar Farahani and Alexander Marinšek},
}

@book{lea2020iot,
  title={IoT and Edge Computing for Architects: Implementing edge and IoT systems from sensors to clouds with communication systems, analytics, and security},
  author={Lea, Perry},
  year={2020},
  publisher={Packt Publishing Ltd}
}

@ARTICLE{894385,
  author={Hubaux, J.-P. and Gross, T. and Le Boudec, J.-Y. and Vetterli, M.},
  journal={IEEE Communications Magazine}, 
  title={Toward self-organized mobile ad hoc networks: the terminodes project}, 
  year={2001},
  volume={39},
  number={1},
  pages={118-124},
  doi={10.1109/35.894385}}

@ARTICLE{8486685,  author={Han, Di and Chen, Wei and Fang, Yuguang},  journal={IEEE Wireless Communications Letters},   title={A Dynamic Pricing Strategy for Vehicle Assisted Mobile Edge Computing Systems},   year={2019},  volume={8},  number={2},  pages={420-423},  doi={10.1109/LWC.2018.2874635}}

@ARTICLE{9293092,
  author={Mai, Tianle and Yao, Haipeng and Guo, Song and Liu, Yunjie},
  journal={IEEE Network}, 
  title={In-Network Computing Powered Mobile Edge: Toward High Performance Industrial IoT}, 
  year={2021},
  volume={35},
  number={1},
  pages={289-295},
  doi={10.1109/MNET.021.2000318}}

@ARTICLE{7931566,  author={Taleb, Tarik and Samdanis, Konstantinos and Mada, Badr and Flinck, Hannu and Dutta, Sunny and Sabella, Dario},  journal={IEEE Communications Surveys   Tutorials},   title={On Multi-Access Edge Computing: A Survey of the Emerging 5G Network Edge Cloud Architecture and Orchestration},   year={2017},  volume={19},  number={3},  pages={1657-1681},  doi={10.1109/COMST.2017.2705720}}

@article{Wang2016DefinitionAC,
  title={Definition and Categorization of Dew Computing},
  author={Yingwei Wang},
  journal={Open J. Cloud Comput.},
  year={2016},
  volume={3},
  pages={1-7}
}

@inproceedings{10.11453359993.3366649,
author = {Albalawi, Abdulazaz Ali and Chakraborti, Asit and Westphal, Cedric and Kutscher, Dirk and He, Jeffrey and Hoole, Quinton},
title = {INCA: An Architecture for In-Network Computing},
year = {2019},
isbn = {9781450370004},
publisher = {Association for Computing Machinery},
address = {New York, NY, USA},
booktitle = {Proceedings of the 1st ACM CoNEXT Workshop on Emerging In-Network Computing Paradigms},
pages = {56–62},
numpages = {7},
series = {ENCP '19}
}

@inproceedings{10.11453445814.3446760,
    author = {Bl\"{o}cher, Marcel and Wang, Lin and Eugster, Patrick and Schmidt, Max},
    title = {Switches for HIRE: Resource Scheduling for Data Center in-Network Computing},
    year = {2021},
    isbn = {9781450383172},
    publisher = {Association for Computing Machinery},
    address = {New York, NY, USA},
    booktitle = {Proceedings of the 26th ACM International Conference on Architectural Support for Programming Languages and Operating Systems},
    pages = {268–285},
    numpages = {18},
    series = {ASPLOS 2021}
    }

@inproceedings{10.11453386367.3431302,
author = {Vaucher, S\'{e}bastien and Yazdani, Niloofar and Felber, Pascal and Lucani, Daniel E. and Schiavoni, Valerio},
title = {ZipLine: In-Network Compression at Line Speed},
year = {2020},
isbn = {9781450379489},
publisher = {Association for Computing Machinery},
address = {New York, NY, USA},
booktitle = {Proceedings of the 16th International Conference on Emerging Networking EXperiments and Technologies},
pages = {399–405},
numpages = {7},
series = {CoNEXT '20}
}

@misc{kim2020unleashing,
      title={Unleashing In-network Computing on Scientific Workloads}, 
      author={Daehyeok Kim and Ankush Jain and Zaoxing Liu and George Amvrosiadis and Damian Hazen and Bradley Settlemyer and Vyas Sekar},
      year={2020},
      eprint={2009.02457},
      archivePrefix={arXiv},
      primaryClass={cs.NI}}

@inproceedings{ACTIVE,
  title={Toward an active network architecture},
  author={Tennenhouse, David L and Wetherall, David J},
  booktitle={Multimedia Computing and Networking 1996},
  volume={2667},
  pages={2--16},
  year={1996},
  organization={International Society for Optics and Photonics}}

@INPROCEEDINGS{8756899,
  author={Tagami, Atsushi and Ueda, Kazuaki and Lukita, Rikisenia and De Benedetto, Jacopo and Arumaithurai, Mayutan and Rossi, Giulio and Detti, Andrea and Hasegawa, Toru},
  booktitle={2019 IEEE International Conference on Communications Workshops (ICC Workshops)}, 
  title={Tile-Based Panoramic Live Video Streaming on ICN}, 
  year={2019},
  volume={},
  number={},
  pages={1-6}}

@ARTICLE{8856241,
  author={Amadeo, Marica and Campolo, Claudia and Ruggeri, Giuseppe and Molinaro, Antonella and Iera, Antonio},
  journal={IEEE Transactions on Network and Service Management}, 
  title={SDN-Managed Provisioning of Named Computing Services in Edge Infrastructures}, 
  year={2019},
  volume={16},
  number={4},
  pages={1464-1478},
  doi={10.1109/TNSM.2019.2945497}}

@ARTICLE{8539024,
  author={Krol, Michal and Marxer, Claudio and Grewe, Dennis and Psaras, Ioannis and Tschudin, Christian},
  journal={IEEE Communications Magazine}, 
  title={Open Security Issues for Edge Named Function Environments}, 
  year={2018},
  volume={56},
  number={11},
  pages={69-75},
  doi={10.1109/MCOM.2018.1701117}}

@ARTICLE{7883826,
  author={Wang, Shuo and Zhang, Xing and Zhang, Yan and Wang, Lin and Yang, Juwo and Wang, Wenbo},
  journal={IEEE Access}, 
  title={A Survey on Mobile Edge Networks: Convergence of Computing, Caching and Communications}, 
  year={2017},
  volume={5},
  number={},
  pages={6757-6779},
  doi={10.1109/ACCESS.2017.2685434}}

@inproceedings{10.1145/3152434.3152461,
author = {Sapio, Amedeo and Abdelaziz, Ibrahim and Aldilaijan, Abdulla and Canini, Marco and Kalnis, Panos},
title = {In-Network Computation is a Dumb Idea Whose Time Has Come},
year = {2017},
publisher = {Association for Computing Machinery},
address = {New York, NY, USA},
booktitle = {Proceedings of the 16th ACM Workshop on Hot Topics in Networks},
pages = {150–156},
numpages = {7},
location = {Palo Alto, CA, USA},
series = {HotNets-XVI}
}

@article{bosshart2014p4,
  title={P4: Programming protocol-independent packet processors},
  author={Bosshart, Pat and Daly, Dan and Gibb, Glen and Izzard, Martin and McKeown, Nick and Rexford, Jennifer and Schlesinger, Cole and Talayco, Dan and Vahdat, Amin and Varghese, George and others},
  journal={ACM SIGCOMM Computer Communication Review},
  volume={44},
  number={3},
  pages={87--95},
  year={2014},
  publisher={ACM New York, NY, USA}
}

@article{yang2019switchagg,
  title={SwitchAgg: A further step towards in-network computation},
  author={Yang, Fan and Wang, Zhan and Ma, Xiaoxiao and Yuan, Guojun and An, Xuejun},
  journal={arXiv preprint arXiv:1904.04024},
  year={2019}
}

@inproceedings{jin2017netcache,
  title={Netcache: Balancing key-value stores with fast in-network caching},
  author={Jin, Xin and Li, Xiaozhou and Zhang, Haoyu and Soul{\'e}, Robert and Lee, Jeongkeun and Foster, Nate and Kim, Changhoon and Stoica, Ion},
  booktitle={Proceedings of the 26th Symposium on Operating Systems Principles},
  pages={121--136},
  year={2017}
}

@article{sapio2019scaling,
  title={Scaling distributed machine learning with in-network aggregation},
  author={Sapio, Amedeo and Canini, Marco and Ho, Chen-Yu and Nelson, Jacob and Kalnis, Panos and Kim, Changhoon and Krishnamurthy, Arvind and Moshref, Masoud and Ports, Dan RK and Richt{\'a}rik, Peter},
  journal={arXiv preprint arXiv:1903.06701},
  year={2019}
}

@INPROCEEDINGS{9350853,  author={Huang, Shuokang and Chen, Ruibiao and Li, Yu and Zhang, Meimei and Lei, Kai and Xu, Ting and Yu, Xiquan},  booktitle={2020 3rd International Conference on Hot Information-Centric Networking (HotICN)},   title={Intelligent Eco Networking (IEN) III: A Shared In-network Computing Infrastructure towards Future Internet},   year={2020},  volume={},  number={},  pages={47-52},  doi={10.1109/HotICN50779.2020.9350853}}

@inproceedings{da2018extern,
  title={Extern objects in P4: An ROHC header compression scheme case study},
  author={da Silva, Jeferson Santiago and Boyer, Fran{\c{c}}ois-Raymond and Chiquette, Laurent-Olivier and Langlois, JM Pierre},
  booktitle={2018 4th IEEE Conference on Network Softwarization and Workshops (NetSoft)},
  pages={517--522},
  year={2018},
  organization={IEEE}
}

@inproceedings{scholz2019cryptographic,
  title={Cryptographic hashing in p4 data planes},
  author={Scholz, Dominik and Oeldemann, Andreas and Geyer, Fabien and Gallenm{\"u}ller, Sebastian and Stubbe, Henning and Wild, Thomas and Herkersdorf, Andreas and Carle, Georg},
  booktitle={2019 ACM/IEEE Symposium on Architectures for Networking and Communications Systems (ANCS)},
  pages={1--6},
  year={2019},
  organization={IEEE}
}

@inproceedings{laki2020price,
  title={The Price for Asynchronous Execution of Extern Functions in Programmable Software Data Planes},
  author={Laki, S{\'a}ndor and Horp{\'a}csi, D{\'a}niel and Voros, Peter and Tejfel, M{\'a}t{\'e} and Hudoba, P{\'e}ter and Pongracz, Gergely and Molnar, Laszlo},
  booktitle={2020 23rd Conference on Innovation in Clouds, Internet and Networks and Workshops (ICIN)},
  pages={23--28},
  year={2020},
  organization={IEEE}
}

@inproceedings{horpacsi2019asynchronous,
  title={Asynchronous extern functions in programmable software data planes},
  author={Horp{\'a}csi, D{\'a}niel and Laki, S{\'a}ndor and V{\"o}r{\"o}s, P{\'e}ter and Tejfel, M{\'a}t{\'e} and Pongr{\'a}cz, Gergely and Moln{\'a}r, L{\'a}szl{\'o}},
  booktitle={2019 ACM/IEEE Symposium on Architectures for Networking and Communications Systems (ANCS)},
  pages={1--2},
  year={2019},
  organization={IEEE}
}

@inproceedings{liu2016one,
  title={One sketch to rule them all: Rethinking network flow monitoring with univmon},
  author={Liu, Zaoxing and Manousis, Antonis and Vorsanger, Gregory and Sekar, Vyas and Braverman, Vladimir},
  booktitle={Proceedings of the 2016 ACM SIGCOMM Conference},
  pages={101--114},
  year={2016}
}

@inproceedings{miao2017silkroad,
  title={Silkroad: Making stateful layer-4 load balancing fast and cheap using switching asics},
  author={Miao, Rui and Zeng, Hongyi and Kim, Changhoon and Lee, Jeongkeun and Yu, Minlan},
  booktitle={Proceedings of the Conference of the ACM Special Interest Group on Data Communication},
  pages={15--28},
  year={2017}
}

@inproceedings{narayana2017language,
  title={Language-directed hardware design for network performance monitoring},
  author={Narayana, Srinivas and Sivaraman, Anirudh and Nathan, Vikram and Goyal, Prateesh and Arun, Venkat and Alizadeh, Mohammad and Jeyakumar, Vimalkumar and Kim, Changhoon},
  booktitle={Proceedings of the Conference of the ACM Special Interest Group on Data Communication},
  pages={85--98},
  year={2017}
}

@article{mckeown2008openflow,
  title={OpenFlow: enabling innovation in campus networks},
  author={McKeown, Nick and Anderson, Tom and Balakrishnan, Hari and Parulkar, Guru and Peterson, Larry and Rexford, Jennifer and Shenker, Scott and Turner, Jonathan},
  journal={ACM SIGCOMM computer communication review},
  volume={38},
  number={2},
  pages={69--74},
  year={2008},
  publisher={ACM New York, NY, USA}
}

@inproceedings{pongracz2013cheap,
  title={Cheap silicon: A myth or reality? Picking the right data plane hardware for software defined networking},
  author={Pongr{\'a}cz, Gergely and Moln{\'a}r, L{\'a}szl{\'o} and Kis, Zolt{\'a}n Lajos and Tur{\'a}nyi, Zolt{\'a}n},
  booktitle={Proceedings of the second ACM SIGCOMM workshop on Hot topics in software defined networking},
  pages={103--108},
  year={2013}
}

@inproceedings{tokusashi2019case,
  title={The case for in-network computing on demand},
  author={Tokusashi, Yuta and Dang, Huynh Tu and Pedone, Fernando and Soul{\'e}, Robert and Zilberman, Noa},
  booktitle={Proceedings of the Fourteenth EuroSys Conference 2019},
  pages={1--16},
  year={2019}
}

@inproceedings{ruth2018towards,
  title={Towards in-network industrial feedback control},
  author={R{\"u}th, Jan and Glebke, Ren{\'e} and Wehrle, Klaus and Causevic, Vedad and Hirche, Sandra},
  booktitle={Proceedings of the 2018 Morning Workshop on In-Network Computing},
  pages={14--19},
  year={2018}
}

@article{li2021advancing,
  title={Advancing Software-Defined Service-Centric Networking Toward In-Network Intelligence},
  author={Li, Xiaolu and Xie, Renchao and Yu, F Richard and Huang, Tao and Liu, Yunjie},
  journal={IEEE Network},
  year={2021},
  publisher={IEEE}
}

@article{bosshart2013forwarding,
  title={Forwarding metamorphosis: Fast programmable match-action processing in hardware for SDN},
  author={Bosshart, Pat and Gibb, Glen and Kim, Hun-Seok and Varghese, George and McKeown, Nick and Izzard, Martin and Mujica, Fernando and Horowitz, Mark},
  journal={ACM SIGCOMM Computer Communication Review},
  volume={43},
  number={4},
  pages={99--110},
  year={2013},
  publisher={ACM New York, NY, USA}
}

@inproceedings{song2013protocol,
  title={Protocol-oblivious forwarding: Unleash the power of SDN through a future-proof forwarding plane},
  author={Song, Haoyu},
  booktitle={Proceedings of the second ACM SIGCOMM workshop on Hot topics in software defined networking},
  pages={127--132},
  year={2013}
}

@article{da2017data,
  title={Data plane programmability beyond openflow: Opportunities and challenges for network and service operations and management},
  author={da Costa Cordeiro, Weverton Luis and Marques, Jonatas Adilson and Gaspary, Luciano Paschoal},
  journal={Journal of Network and Systems Management},
  volume={25},
  number={4},
  pages={784--818},
  year={2017},
  publisher={Springer}
}

@inproceedings{woodruff2019p4dns,
  title={P4DNS: in-network DNS},
  author={Woodruff, Jackson and Ramanujam, Murali and Zilberman, Noa},
  booktitle={2019 ACM/IEEE Symposium on Architectures for Networking and Communications Systems (ANCS)},
  pages={1--6},
  year={2019},
  organization={IEEE}
}

@inproceedings{grigoryan2019iload,
  title={iload: In-network load balancing with programmable data plane},
  author={Grigoryan, Garegin and Liu, Yaoqing and Kwon, Minseok},
  booktitle={Proceedings of the 15th International Conference on emerging Networking EXperiments and Technologies},
  pages={17--19},
  year={2019}
}

@inproceedings{liu2017incbricks,
  title={Incbricks: Toward in-network computation with an in-network cache},
  author={Liu, Ming and Luo, Liang and Nelson, Jacob and Ceze, Luis and Krishnamurthy, Arvind and Atreya, Kishore},
  booktitle={Proceedings of the Twenty-Second International Conference on Architectural Support for Programming Languages and Operating Systems},
  pages={795--809},
  year={2017}
}

@inproceedings{Coicn,
author = {Kr\'{o}l, Micha\l{} and Nicolaescu, Adrian-Cristian and Re\~{n}\'{e}, Sergi and Ascigil, Onur and Psaras, Ioannis and Oran, David and Kutscher, Dirk},
title = {Computation Offloading with ICN},
year = {2018},
isbn = {9781450359597},
publisher = {Association for Computing Machinery},
address = {New York, NY, USA},
doi = {10.1145/3267955.3269009},
booktitle = {Proceedings of the 5th ACM Conference on Information-Centric Networking},
pages = {220–221},
numpages = {2},
location = {Boston, Massachusetts},
series = {ICN '18}
}

@inproceedings{cloud1,
author = {Steiner, Moritz and Gaglianello, Bob Gaglianello and Gurbani, Vijay and Hilt, Volker and Roome, W.D. and Scharf, Michael and Voith, Thomas},
title = {Network-Aware Service Placement in a Distributed Cloud Environment},
year = {2012},
isbn = {9781450314190},
publisher = {Association for Computing Machinery},
address = {New York, NY, USA},
url = {https://doi.org/10.1145/2342356.2342366},
doi = {10.1145/2342356.2342366},
booktitle = {Proceedings of the ACM SIGCOMM 2012 Conference on Applications, Technologies, Architectures, and Protocols for Computer Communication},
pages = {73–74},
numpages = {2},
location = {Helsinki, Finland},
series = {SIGCOMM '12}
}

@INPROCEEDINGS{LightNF,  author={Chen, Xiang and Huang, Qun and Wang, Peiqiao and Meng, Zili and Liu, Hongyan and Chen, Yuxin and Zhang, Dong and Zhou, Haifeng and Zhou, Boyang and Wu, Chunming},  booktitle={2021 IEEE/ACM 29th International Symposium on Quality of Service (IWQOS)},   title={LightNF: Simplifying Network Function Offloading in Programmable Networks},   year={2021},  volume={},  number={},  pages={1-10},  doi={10.1109/IWQOS52092.2021.9521329}}

@inproceedings{10.1145/3125719.3125727,
author = {Kr\'{o}l, Micha\l{} and Psaras, Ioannis},
title = {NFaaS: Named Function as a Service},
year = {2017},
isbn = {9781450351225},
publisher = {Association for Computing Machinery},
address = {New York, NY, USA},
doi = {10.1145/3125719.3125727},
booktitle = {Proceedings of the 4th ACM Conference on Information-Centric Networking},
pages = {134–144},
numpages = {11},
series = {ICN '17}
}

@article{NDNsurvey,
title = {Named Data Networking: A survey},
journal = {Computer Science Review},
volume = {19},
pages = {15-55},
year = {2016},
issn = {1574-0137},
doi = {10.1016/j.cosrev.2016.01.001},
author = {Divya Saxena and Vaskar Raychoudhury and Neeraj Suri and Christian Becker and Jiannong Cao},
}

@ARTICLE{NDNwireless,
  author={Tariq, Asadullah and Rehman, Rana Asif and Kim, Byung-Seo},
  journal={IEEE Communications Surveys   Tutorials}, 
  title={Forwarding Strategies in NDN-Based Wireless Networks: A Survey}, 
  year={2020},
  volume={22},
  number={1},
  pages={68-95},
  doi={10.1109/COMST.2019.2935795}}

@ARTICLE{INC6G,  author={Hu, Ning and Tian, Zhihong and Du, Xiaojiang and Guizani, Mohsen},  journal={IEEE Transactions on Green Communications and Networking},   title={An Energy-Efficient In-Network Computing Paradigm for 6G},   year={2021},  volume={},  number={},  pages={1-1},  doi={10.1109/TGCN.2021.3099804}}

@ARTICLE{FASE,  author={Lee, Jaewook and Ko, Haneul and Lee, Hochan and Pack, Sangheon},  journal={IEEE Access},   title={Flow-Aware Service Function Embedding Algorithm in Programmable Data Plane},   year={2021},  volume={9},  number={},  pages={6113-6121},  doi={10.1109/ACCESS.2020.3048421}}

@INPROCEEDINGS{P4SC,  author={Chen, Xiang and Zhang, Dong and Wang, Xiaojun and Zhu, Kai and Zhou, Haifeng},  booktitle={2019 IFIP/IEEE Symposium on Integrated Network and Service Management (IM)},   title={P4SC: Towards High-Performance Service Function Chain Implementation on the P4-Capable Device},   year={2019},  volume={},  number={},  pages={1-9},  doi={}}

@inproceedings{RESFC,
  title={Resource-efficient service function chaining in programmable data plane},
  author={Lee, H and Lee, J and Ko, H and Pack, S},
  booktitle={Proc. EuroP4},
  year={2019}}

@inproceedings{AccSFC,
author = {Wu, Dingming and Chen, Ang and Ng, T. S. Eugene and Wang, Guohui and Wang, Haiyong},
title = {Accelerated Service Chaining on a Single Switch ASIC},
year = {2019},
isbn = {9781450370202},
publisher = {Association for Computing Machinery},
address = {New York, NY, USA},
doi = {10.1145/3365609.3365849},
booktitle = {Proceedings of the 18th ACM Workshop on Hot Topics in Networks},
pages = {141–149},
numpages = {9},
location = {Princeton, NJ, USA},
series = {HotNets '19}
}

@INPROCEEDINGS{PIAFFE,  author={Mafioletti, Diego Rossi and Dominicini, Cristina Klippel and Martinello, Magnos and Ribeiro, Moises R. N. and Villaça, Rodolfo da Silva},  booktitle={ICC 2020 - 2020 IEEE International Conference on Communications (ICC)},   title={PIaFFE: A Place-as-you-go In-network Framework for Flexible Embedding of VNFs},   year={2020},  volume={},  number={},  pages={1-6},  doi={10.1109/ICC40277.2020.9149240}}

@inproceedings{NFCC,
  title={Named functions and cached computations},
  author={Tschudin, Christian and Sifalakis, Manolis},
  booktitle={2014 IEEE 11th consumer communications and networking conference (CCNC)},
  pages={851--857},
  year={2014},
  organization={IEEE}
}

@inproceedings{RICE,
  title={Rice: Remote method invocation in icn},
  author={Kr{\'o}l, Micha{\l} and Habak, Karim and Oran, David and Kutscher, Dirk and Psaras, Ioannis},
  booktitle={Proceedings of the 5th ACM Conference on Information-Centric Networking},
  pages={1--11},
  year={2018}
}

@inproceedings{placeDelayINC,
  title={Optimal Placement of Delay-constrained In-Network Computing Tasks at the Edge with Minimum Data Exchange},
  author={Lia, Gianmarco and Amadeo, Marica and Campolo, Claudia and Ruggeri, Giuseppe and Molinaro, Antonella},
  booktitle={2021 IEEE 4th 5G World Forum (5GWF)},
  pages={481--486},
  year={2021},
  organization={IEEE}
}

@article{distributedINC,
  title={A model for distributed in-network and near-edge computing with heterogeneous hardware},
  author={Cooke, Ryan A and Fahmy, Suhaib A},
  journal={Future Generation Computer Systems},
  volume={105},
  pages={395--409},
  year={2020},
  publisher={Elsevier}}

@article{placementSurvey,
  title={Survey on Placement Methods in the Edge and Beyond},
  author={Sonkoly, Bal{\'a}zs and Czentye, J{\'a}nos and Szalay, M{\'a}rk and N{\'e}meth, Bal{\'a}zs and Toka, L{\'a}szl{\'o}},
  journal={IEEE Communications Surveys \& Tutorials},
  year={2021},
  publisher={IEEE}
}

@article{VNFMicro,
  title={Re-architecting NFV ecosystem with microservices: State of the art and research challenges},
  author={Chowdhury, Shihabur Rahman and Salahuddin, Mohammad A and Limam, Noura and Boutaba, Raouf},
  journal={IEEE Network},
  volume={33},
  number={3},
  pages={168--176},
  year={2019},
  publisher={IEEE}
}

@ARTICLE{TSEC,  author={Chen, Liqiong and Guo, Kun and Fan, Guoqing and Wang, Can and Song, Shilong},  journal={IEEE Access},   title={Resource Constrained Profit Optimization Method for Task Scheduling in Edge Cloud},   year={2020},  volume={8},  number={},  pages={118638-118652},  doi={10.1109/ACCESS.2020.3000985}}

@article{TSVNF,
  title={Task scheduling for edge computing with agile VNFs on-demand service model toward 5G and beyond},
  author={Tseng, Chia-Wei and Tseng, Fan-Hsun and Yang, Yao-Tsung and Liu, Chien-Chang and Chou, Li-Der},
  journal={Wireless Communications and Mobile Computing},
  volume={2018},
  year={2018},
  publisher={Hindawi}
}

@inproceedings{HealthEdge,
  title={Healthedge: Task scheduling for edge computing with health emergency and human behavior consideration in smart homes},
  author={Wang, Haoyu and Gong, Jiaqi and Zhuang, Yan and Shen, Haiying and Lach, John},
  booktitle={2017 IEEE International Conference on Big Data (Big Data)},
  pages={1213--1222},
  year={2017},
  organization={IEEE}
}

@inproceedings{ENTICE,
  title={The ENTICE approach to decompose monolithic services into microservices},
  author={Kecskemeti, Gabor and Marosi, Attila Csaba and Kertesz, Attila},
  booktitle={2016 International Conference on High Performance Computing \& Simulation (HPCS)},
  pages={591--596},
  year={2016},
  organization={IEEE}
}

@inproceedings{VirtualizationPerformance,
  title={Performance analysis of virtual machines and containers in cloud computing},
  author={Barik, Rabindra K and Lenka, Rakesh K and Rao, K Rahul and Ghose, Devam},
  booktitle={2016 international conference on computing, communication and automation (iccca)},
  pages={1204--1210},
  year={2016},
  organization={IEEE}
}

@inproceedings{40,
  title={Security-as-a-service for microservices-based cloud applications},
  author={Sun, Yuqiong and Nanda, Susanta and Jaeger, Trent},
  booktitle={2015 IEEE 7th International Conference on Cloud Computing Technology and Science (CloudCom)},
  pages={50--57},
  year={2015},
  organization={IEEE}
}

@inproceedings{46,
  title={Container and microservice driven design for cloud infrastructure devops},
  author={Kang, Hui and Le, Michael and Tao, Shu},
  booktitle={2016 IEEE International Conference on Cloud Engineering (IC2E)},
  pages={202--211},
  year={2016},
  organization={IEEE}
}

@inproceedings{Dan2019,
  title={When Should the Network Be the Computers},
  author={Dan R. K. Ports and Jacob Nelson},
  booktitle={HotOS '19: Proceedings of the Workshop on Hot Topics in Operating Systems},
  pages={209-215},
  year={2019}
}

@inproceedings{anwer2015programming,
  title={Programming slick network functions},
  author={Anwer, Bilal and Benson, Theophilus and Feamster, Nick and Levin, Dave},
  booktitle={Proceedings of the 1st acm sigcomm symposium on software defined networking research},
  pages={1--13},
  year={2015}
}

@inproceedings{mustard2019jumpgate,
  title={Jumpgate: In-network processing as a service for data analytics},
  author={Mustard, Craig and Ruffy, Fabian and Gakhokidze, Anny and Beschastnikh, Ivan and Fedorova, Alexandra},
  booktitle={11th $\{$USENIX$\}$ Workshop on Hot Topics in Cloud Computing (HotCloud 19)},
  year={2019}
}

@article{blocher2021holistic,
  title={Holistic Runtime Scheduling for the Distributed Computing Landscape},
  author={Bl{\"o}cher, Marcel},
  year={2021},
  publisher={Technische Universit{\"a}t Darmstadt}
}

@inproceedings{mai2014netagg,
  title={Netagg: Using middleboxes for application-specific on-path aggregation in data centres},
  author={Mai, Luo and Rupprecht, Lukas and Alim, Abdul and Costa, Paolo and Migliavacca, Matteo and Pietzuch, Peter and Wolf, Alexander L},
  booktitle={Proceedings of the 10th ACM International on Conference on emerging Networking Experiments and Technologies},
  pages={249--262},
  year={2014}
}

@inproceedings{tokusashi2018lake,
  title={LaKe: the power of in-network computing},
  author={Tokusashi, Yuta and Matsutani, Hiroki and Zilberman, Noa},
  booktitle={2018 International Conference on ReConFigurable Computing and FPGAs (ReConFig)},
  pages={1--8},
  year={2018},
  organization={IEEE}
}

@inproceedings{ports2019should,
  title={When should the network be the computer?},
  author={Ports, Dan RK and Nelson, Jacob},
  booktitle={Proceedings of the Workshop on Hot Topics in Operating Systems},
  pages={209--215},
  year={2019}
}

@inproceedings{istvan2016consensus,
  title={Consensus in a box: Inexpensive coordination in hardware},
  author={Istv{\'a}n, Zsolt and Sidler, David and Alonso, Gustavo and Vukolic, Marko},
  booktitle={13th $\{$USENIX$\}$ Symposium on Networked Systems Design and Implementation ($\{$NSDI$\}$ 16)},
  pages={425--438},
  year={2016}
}

@inproceedings{zhang2014smartswitch,
  title={Smartswitch: Blurring the line between network infrastructure \& cloud applications},
  author={Zhang, Wei and Wood, Timothy and Ramakrishnan, KK and Hwang, Jinho},
  booktitle={6th $\{$USENIX$\}$ Workshop on Hot Topics in Cloud Computing (HotCloud 14)},
  year={2014}
}

@inproceedings{alim2016flick,
  title={$\{$FLICK$\}$: Developing and Running Application-Specific Network Services},
  author={Alim, Abdul and Clegg, Richard G and Mai, Luo and Rupprecht, Lukas and Seckler, Eric and Costa, Paolo and Pietzuch, Peter and Wolf, Alexander L and Sultana, Nik and Crowcroft, Jon and others},
  booktitle={2016 $\{$USENIX$\}$ Annual Technical Conference ($\{$USENIX$\}$$\{$ATC$\}$ 16)},
  pages={1--14},
  year={2016}
}

@ARTICLE{7879258,  author={Mach, Pavel and Becvar, Zdenek},  journal={IEEE Communications Surveys   Tutorials},   title={Mobile Edge Computing: A Survey on Architecture and Computation Offloading},   year={2017},  volume={19},  number={3},  pages={1628-1656},  doi={10.1109/COMST.2017.2682318}}

@ARTICLE{9519636,  author={Luo, Quyuan and Hu, Shihong and Li, Changle and Li, Guanghui and Shi, Weisong},  journal={IEEE Communications Surveys   Tutorials},   title={Resource Scheduling in Edge Computing: A Survey},   year={2021},  volume={23},  number={4},  pages={2131-2165},  doi={10.1109/COMST.2021.3106401}}

@ARTICLE{8896954,  author={Mehrabi, Mahshid and You, Dongho and Latzko, Vincent and Salah, Hani and Reisslein, Martin and Fitzek, Frank H. P.},  journal={IEEE Access},   title={Device-Enhanced MEC: Multi-Access Edge Computing (MEC) Aided by End Device Computation and Caching: A Survey},   year={2019},  volume={7},  number={},  pages={166079-166108},  doi={10.1109/ACCESS.2019.2953172}}

@article{10.1145/3362031,
author = {Ren, Ju and Zhang, Deyu and He, Shiwen and Zhang, Yaoxue and Li, Tao},
title = {A Survey on End-Edge-Cloud Orchestrated Network Computing Paradigms: Transparent Computing, Mobile Edge Computing, Fog Computing, and Cloudlet},
year = {2019},
issue_date = {November 2020},
publisher = {Association for Computing Machinery},
address = {New York, NY, USA},
volume = {52},
number = {6},
issn = {0360-0300},
doi = {10.1145/3362031},
journal = {ACM Comput. Surv.},
month = {oct},
articleno = {125},
numpages = {36},
}

@misc{NoaZilberman,
  title = "In network Computing",
  published  = "online article",
  organization  = "Noa Zilberman",
  month         = April,
  year          = "2019"
}

@article{tang2020communication,
  title={Communication, computation, and caching resource sharing for the internet of things},
  author={Tang, Ming and Gao, Lin and Huang, Jianwei},
  journal={IEEE Communications Magazine},
  volume={58},
  number={4},
  pages={75--80},
  year={2020},
  publisher={IEEE}
}

@inproceedings{InNet,
  title={In-net: In-network processing for the masses},
  author={Stoenescu, Radu and Olteanu, Vladimir and Popovici, Matei and Ahmed, Mohamed and Martins, Joao and Bifulco, Roberto and Manco, Filipe and Huici, Felipe and Smaragdakis, Georgios and Handley, Mark and others},
  booktitle={Proceedings of the Tenth European Conference on Computer Systems},
  pages={1--15},
  year={2015}
}

@inproceedings{sifalakis2014information,
  title={An information centric network for computing the distribution of computations},
  author={Sifalakis, Manolis and Kohler, Basil and Scherb, Christopher and Tschudin, Christian},
  booktitle={Proceedings of the 1st ACM Conference on Information-Centric Networking},
  pages={137--146},
  year={2014}
}

@inproceedings{scherb2017execution,
  title={Execution state management in named function networking},
  author={Scherb, Christopher and Faludi, Bal{\'a}zs and Tschudin, Christian},
  booktitle={2017 IFIP Networking Conference (IFIP Networking) and Workshops},
  pages={1--6},
  year={2017},
  organization={IEEE}
}

@article{da2020pipelining,
  title={Pipelining for Named Function Networking},
  author={da Costa, Eric},
  year={2020}
}

@article{zhan2020incentive,
  title={An incentive mechanism design for mobile crowdsensing with demand uncertainties},
  author={Zhan, Yufeng and Xia, Yuanqing and Zhang, Jiang and Li, Ting and Wang, Yu},
  journal={Information Sciences},
  volume={528},
  pages={1--16},
  year={2020},
  publisher={Elsevier}
}

@misc{Zilberman,
  author       = "Noa Zilberman",
  title        = "In-Network Computing",
  howpublished = "https://www.sigarch.org/in-network-computing-draft",
  month        = "April",
  year         = "2019",
}

@ARTICLE{8758431,
  author={Zhao, Yongli and Wang, Wei and Li, Yajie and Colman Meixner, Carlos and Tornatore, Massimo and Zhang, Jie},
  journal={IEEE Access}, 
  title={Edge Computing and Networking: A Survey on Infrastructures and Applications}, 
  year={2019},
  volume={7},
  number={},
  pages={101213-101230},
  doi={10.1109/ACCESS.2019.2927538}}

@ARTICLE{8714026,
  author={Luong, Nguyen Cong and Hoang, Dinh Thai and Gong, Shimin and Niyato, Dusit and Wang, Ping and Liang, Ying-Chang and Kim, Dong In},
  journal={IEEE Communications Surveys   Tutorials}, 
  title={Applications of Deep Reinforcement Learning in Communications and Networking: A Survey}, 
  year={2019},
  volume={21},
  number={4},
  pages={3133-3174},
  doi={10.1109/COMST.2019.2916583}}

@article{bittencourt2018internet,
  title={The internet of things, fog and cloud continuum: Integration and challenges},
  author={Bittencourt, Luiz and Immich, Roger and Sakellariou, Rizos and Fonseca, Nelson and Madeira, Edmundo and Curado, Marilia and Villas, Leandro and DaSilva, Luiz and Lee, Craig and Rana, Omer},
  journal={Internet of Things},
  volume={3},
  pages={134--155},
  year={2018},
  publisher={Elsevier}
}

@ARTICLE{9476028,  author={Santos, José and Wauters, Tim and Volckaert, Bruno and De Turck, Filip},  journal={IEEE Communications Surveys   Tutorials},   title={Towards Low-Latency Service Delivery in a Continuum of Virtual Resources: State-of-the-Art and Research Directions},   year={2021},  volume={23},  number={4},  pages={2557-2589},  doi={10.1109/COMST.2021.3095358}}

@ARTICLE{8976180,
  author={Wang, Xiaofei and Han, Yiwen and Leung, Victor C. M. and Niyato, Dusit and Yan, Xueqiang and Chen, Xu},
  journal={IEEE Communications Surveys   Tutorials}, 
  title={Convergence of Edge Computing and Deep Learning: A Comprehensive Survey}, 
  year={2020},
  volume={22},
  number={2},
  pages={869-904},
  doi={10.1109/COMST.2020.2970550}}

@ARTICLE{INC22,  author={Kianpisheh, Somayeh and Taleb, Tarik},  journal={IEEE Communications Surveys & Tutorials},   title={A Survey on In-network Computing: Programmable Data Plane And Technology Specific Applications},   year={2022},  volume={},  number={},  pages={1-1},  doi={10.1109/COMST.2022.3213237}}


%% file: stateofart.bib
@ARTICLE{7543455,  author={Dastjerdi, Amir Vahid and Buyya, Rajkumar},  journal={Computer},   title={Fog Computing: Helping the Internet of Things Realize Its Potential},   year={2016},  volume={49},  number={8},  pages={112-116},  doi={10.1109/MC.2016.245}}

@INPROCEEDINGS{8264868,  author={Scherb, Christopher and Marxer, Claudio and Schnurrenberger, Urs and Tschudin, Christian},  booktitle={2017 IFIP Networking Conference (IFIP Networking) and Workshops},   title={In-network live stream processing with named functions},   year={2017},  volume={},  number={},  pages={1-6},  doi={10.23919/IFIPNetworking.2017.8264868}}

@article{zeng2021guest,
  title={Guest editorial: In-network computing: Emerging trends for the edge-cloud continuum},
  author={Zeng, Deze and Ansari, Nirwan and Montpetit, Marie-Jos{\'e} and Schooler, Eve M and Tarchi, Daniele},
  journal={IEEE Network},
  volume={35},
  number={5},
  pages={12--13},
  year={2021},
  publisher={IEEE}
}

@misc{Tofino,
    author = {{}},
    title = {{Intel® Tofino™
P4-programmable Ethernet switch ASIC that delivers better performance at lower power}},
    howpublished = {\url{https://www.intel.com/content/www/us/en/products/network-io/programmable-ethernet-switch/tofino-series.html}},
    note = {Online; accessed 04 October 2022}}
